\documentclass[12pt]{iopart}

\usepackage{iopams}

\expandafter\let\csname equation*\endcsname\relax

\expandafter\let\csname endequation*\endcsname\relax
\usepackage{comment}
\usepackage{amsmath}
\usepackage{cite} 
\usepackage{tikz}
\usepackage{graphicx}
\usepackage{soul}
\usetikzlibrary{matrix,shapes,arrows,positioning,chains} 
\usepackage{hyperref}
\usepackage{amsmath}
\usepackage{amsfonts}
\usepackage{amssymb,latexsym,mathrsfs}
\usepackage{graphicx} 
\usepackage{float}

\begin{document}

\title[Transition to chaos in extended systems and their quantum impurity models]{Transition to chaos in extended systems and their quantum impurity models}
%\newcommand{\titlename}{Transition to chaos in extended systems and their quantum impurity models}
%\title{\titlename}

\author{Mahaveer Prasad$^1$, Hari Kumar Yadalam$^{1,2,3,4}$, Manas Kulkarni$^1$, Camille Aron$^{2,5}$} 

\address{$^1$ International Centre for Theoretical Sciences, Tata Institute of Fundamental Research, Bengaluru -- 560089, India}
\address{$^2$ Laboratoire de Physique de l'\'Ecole Normale Sup\'erieure, ENS, Universit\'e PSL, CNRS, Sorbonne Universit\'e, Universit\'e Paris Cit\'e, F-75005 Paris,  France \\}
\address{$^3$ Department of Chemistry, University of California, Irvine, CA 92614, USA}
\address{$^4$ Department of Physics and Astronomy, University of California, Irvine, CA 92614, USA} 
\address{$^5$ Institute of Physics, \'Ecole Polytechnique F\'ed\'erale de Lausanne (EPFL), CH-1015 Lausanne, Switzerland}
\ead{aron@ens.fr}

\date{\today}
	
\begin{abstract}
Chaos sets a fundamental limit to quantum-information processing schemes.
We study the onset of chaos in spatially extended quantum many-body systems that are relevant to quantum optical devices.
We consider an extended version of the Tavis-Cummings model on a finite chain.
By studying level-spacing statistics, adjacent gap ratios, and spectral form factors, we observe the transition from integrability to chaos as the hopping between the Tavis-Cummings sites is increased above a finite value.
The results are obtained by means of exact numerical diagonalization which becomes notoriously hard for extended lattice geometries.
In an attempt to circumvent these difficulties, we identify a minimal single-site quantum impurity model that successfully captures the spectral properties of the lattice model. 
This approach is intended to be adaptable to other lattice models with large local Hilbert spaces.
\end{abstract}

%\keywords{Quantum chaos, impurity model}

\maketitle

\section{Introduction}
The field of quantum optics is making remarkable progress towards incorporating more and more controllable quantum degrees of freedom in its 
devices~\cite{Xiang2013Hybrid,Aspelmeyer2014cavity,kurizki2015quantumtech,Noh2016quantum,LEHUR2016manybody,Cottet2017cavityQED,clerk2020hybrid,tangpanitanon2020many, haroche2020cavity,blais2020quantum,Blais2021circuit,hirayama2021hybrid}. 
Studying the onset of chaos in those quantum-information processing schemes is therefore not only of fundamental relevance but it has also become a pressing practical issue~\cite{Vitanov2017simulated,Day2018adiabatic,Day2019manybody,
dey2020emergence,Lbez2021,berke2022transmon}.
Indeed, integrability is a brittle property of rare and specific models. Perhaps with the exception of many-body localized systems, generic perturbations to a spatially extended integrable system are expected to immediately break its integrability, and to bring chaotic dynamics~\cite{Santos2004integrability,Torres2014local,Brenes2018High, Bastianello2019Lack,Brenes2020Eigenstate,znidaric2020weak}. In particular, this scenario is expected in the \emph{thermodynamic} scaling regime, \textit{i.e} whenever the perturbation contributes extensively to the energy.
For applications, it is desirable to rather operate with integrable dynamics to avoid the scrambling of quantum information. This is achieved by working in the so-called \emph{dynamic} scaling regime where the integrability-breaking terms are scaled down adequately as one increases the system size~\cite{bulchandani2021onset} so as to act as irrelevant perturbations from the standpoint of spectral statistics.

Given their Hilbert space which typically grows exponentially with their size, lattice problems are notoriously hard to deal with exact-diagonalization techniques. To evade this difficultly, we seek a minimal quantum impurity model that can reproduce the  spectral features of the lattice problem at a lower computational cost.
In contrast to the conventional practice of defining impurity models in the thermodynamic limit, say for addressing thermodynamic phase transitions, we propose an implementation in the above-mentioned dynamic scaling regime.
The basic intuition behind our impurity modeling is the following.
In the integrable phase of a lattice model composed of integrable unit cells, the local integrals of motion (LIOMs)~\cite{Chandran2015Constructing,ROS2015Integrals,Rademaker2016Explicit,Imbrie2017Locals,huse2014phenomenology} will be localized about the lattice sites. Hence, a smaller scale description in terms of a few lattice sites should suffice to capture those integrable features. 
We stress that it is the dynamic scaling which ensures the proximity of an integrable phase with local conserved charges.
This justifies the idea of considering spectral statistics of unit cells as a local order parameter.
In the chaotic phase of the lattice model, the degrees of freedom are delocalized and the integrability of the corresponding impurity model has to be broken concomitantly.

We examine these ideas in the framework of the Tavis-Cummings lattice (TCL) which is an archetypal model of local quantum degrees of freedom coupled to itinerant photons relevant for numerous experimental platforms~\cite{schmidt2013circuit,zou2014implementation,kurizki2015quantumtech}. It consists of a collection of Tavis-Cummings (TC) models loaded on a tight-binding lattice. 
The TC model has a large local Hilbert space and is known to be integrable.
When loaded on a finite lattice, the integrability is expected to be broken and the chaos to set in at finite values of the hopping amplitude.
As an associated impurity model, we shall consider the single-site TC model driven by a coherent source mimicking the coupling to neighbors and breaking the integrability of the undriven impurity.

After we introduce the TCL and its associated impurity model, we characterize their respective transition from integrability to chaos by means of extensive exact-diagonalization computations. We extract the statistical properties of their spectra and compute their level-spacing distributions, adjacent gap ratios, and spectral form factors. We find the spectral properties of the TCL to transition from Poisson statistics to those of random matrix theory (RMT) as one increases the hopping amplitude.
Remarkably, the spectral form factors are computed from disorder-free models and without averaging over any model parameter. We show that the associated impurity model can successfully reproduce the spectral features of the lattice model and we compute the map between the integrability-breaking parameters of both models.

\begin{figure}[!tbp]
\centering
\includegraphics[scale=0.25]{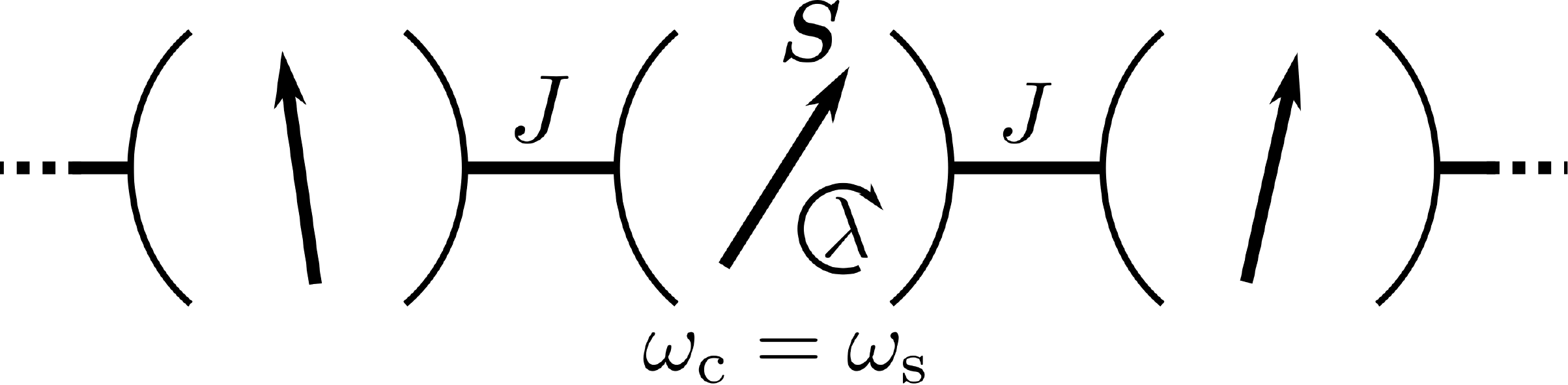}
\caption{Tavis-Cummings lattice (TCL): Tavis-Cummings units hosted on a one-dimensional tight-binding lattice of size $L$ with open boundary conditions.
Each unit features a large spin $S$ coupled to a bosonic mode via the interaction $\lambda$.
$J$ sets the hopping amplitude of the bosons between neighboring units.
See the Hamiltonian in Eq.~(\ref{eq:H_latt}).
}
\label{fig:sch}
\end{figure}

\section{Tavis-Cummings Lattice (TCL)}
The TCL describes an extended array of large quantum spins coupled via photon-mediated interactions~\cite{tavis1968exact,rossini2007mott}. 
We consider the Hamiltonian
\begin{align} 
H = & \sum_{i=1}^L h_i + \sum_{\langle i j \rangle} h_{ij} \,, \label{eq:H_latt} \\
h_{i} = & \, \omega_c a_i^{\dagger}a_i + \omega_s S_i^{z}  
+\frac{\lambda}{\sqrt{S}} \left( a_i^{\dagger} S_i^- +  a_i S_i^+  \right)  \,,
\nonumber \\
h_{ij}= & - \frac{J}{2} \left( a_i^\dagger a_j +  a_j^\dagger a_i \right) \,, \nonumber
\end{align}
where individual Tavis-Cummings (TC) models, with Hamiltonians $h_i$, are loaded on a one-dimensional tight-binding lattice with $L$ sites and open boundary conditions.
$a_i$ ($a_i^{\dagger}$) is the bosonic annihilation (creation) operator of the cavity mode at site $i$ with energy $\omega_c$. $S_i^{\alpha}$, $\alpha=x,y,z$, are the spin angular momentum operators built from the totally symmetric representation of $S$ identical two-level systems with energy splitting $\omega_s$. 
Throughout the paper, we consider the resonant regime $\omega_c=\omega_s = \omega_0$ and set the unit of energy  $\omega_0 = 1$.
$\lambda$ sets the interaction strength between spins and cavity modes. 
$h_{ij}$ introduces coherent hopping amplitude $J> 0$ between the nearest-neighbor cavity modes.
In the atomic limit, $J=0$, one recovers the physics of the single-site TC model: in the $S\to\infty$ limit, $\lambda > 1$ drives a spontaneous $U(1)$ symmetry-breaking 
quantum phase transition between a normal and a superradiant 
phase~\cite{Hepp1973on,Hioe1973phase,Wang1973phase,
Hepp1973Equilibrium,roses2020dicke}.
Notably, the TC model is integrable on both sides of the phase 
transition~\cite{LEWENKOPF1991Level,emary2003chaos1,Buijsman2017Nonergodicity}.
The hopping $J> 0$ demotes the local $U(1)$ symmetry of the TC model to a global $U(1)$ symmetry in the TCL model corresponding to the conservation of the total number of excitations.
The normal phase of the TC model extends in the $J$--$\lambda$ plane of the phase diagram of the TCL model.
Importantly, $J$ acts as an integrability-breaking parameter. More precisely, for finite size lattices, the integrable character of the TC model is expected to be robust until a finite value of $J$ which rapidly vanishes as the size of the lattice and of the local Hilbert space, respectively $L$ and $S$, are increased.
In practice, we explore the onset of chaos by working at $L = 3$ cavities, which is experimentally feasible and relevant. Furthermore, while that the spin-boson scaling factor $1/\sqrt{S}$ in (\ref{eq:H_latt}) is conventionally introduced  to ensure non-trivial thermodynamics in the $S \to \infty$ limit, we shall see below that a proper dynamic scaling, in the sense of Bulchandani, Huse and Gopalakrishnan in Ref.~\cite{bulchandani2021onset}, requires the extra rescaling $J \to J/S^{1/4}$.

\section{Impurity model}
Let us now introduce the impurity model associated to the above lattice model. It is given by the Hamiltonian
\begin{align}
\label{eq:H_imp}
H_{\rm imp} = & \, \omega_c a^{\dagger}a + \omega_s S^{z}  
+\frac{\lambda}{\sqrt{S}} \left( a^{\dagger} S^- +  a S^+  \right) -  \mu \sqrt{S} \left( a+  a^\dagger \right)\,.
\end{align}
It corresponds to a single-site TC model with an additional drive term controlled by the parameter $\mu$.
Similarly to the lattice model, we set $\omega_c = \omega_s = \omega_0$. Note that the parameter $\omega_0$ of the impurity can in principle be different from the one of the lattice model but, for simplicity, we also set it as the unit of energy.
Similarly to the integrability-breaking parameter $J/\lambda$ that lifts the local $U(1)$ symmetry of the lattice model to a global $U(1)$ symmetry, the impurity drive $\mu \neq 0$ explicitly breaks the $U(1)$ symmetry as well as the integrability of the TC model. {Note that the scaling factors of the spin-boson interaction and the drive in $H_{\rm imp}$, respectively $1/\sqrt{S}$ and $\sqrt{S}$, ensure non-trivial thermodynamics in the $S \to \infty$ limit. While we follow that convention, we shall see later that a proper dynamic scaling, in the sense of Bulchandani, Huse and Gopalakrishnan in Ref.~\cite{bulchandani2021onset}, requires rescaling the drive $\mu$ with a factor  $1/S^{1/4}$ rather than $\sqrt{S}$.}
Similar single-site models have been used in the literature to study the stability of the superradiant phase-transition and the onset of quantum chaos~\cite{provost1976lack,corps2022chaos}.
Intuitively, the drive term in $H_{\rm imp}$ can be seen as mimicking the hopping from the rest of the lattice on the impurity site. In that view, $\mu$ is expected to depend on the size and the precise geometry of the lattice, and it vanishes in the atomic limit $J \to 0$.
We motivate our choice of impurity model in Eq.~(\ref{eq:H_imp}) from the fact that in the dynamic scaling regime, where the integrability-breaking parameter $J/\lambda$ is small, it can be derived from a lattice model with a large coordination number using a standard mean-field approach. We refer the reader to the~\ref{lattice-impurity} for a detailed presentation of this construction.
Notably, we found the classical version of the driven impurity model to unambiguously exhibit chaotic dynamics for intermediate values of $\mu$.
We refer the reader to the~\ref{classical} for a detailed analysis.
The conjecture by Bohigas, Giannoni and Schmit (BHS)~\cite{bohigas1984characterization} states that those Hamiltonians with a chaotic classical limit have spectra whose statistical features are governed by RMT. As a consequence, we expect the quantum impurity model in Eq.~(\ref{eq:H_imp}) to exhibit RMT features.

\section{Spectral properties}
We analyze the statistical properties of the eigenvalues $\{E_{n}\}$ of both the lattice Hamiltonian $H$ in Eq.~(\ref{eq:H_latt}) and the impurity Hamiltonian $H_{\rm imp}$ in Eq.~(\ref{eq:H_imp}) by means of exact diagonalization.
Given the spatial reflection symmetry and the $U(1)$ symmetry of the finite lattice model, we choose to compute the spectral statistics from the reflection-symmetric sector with a fixed number of excitations, labeled by the quantum number $N_{\rm ex} = 36$.
Therefore, the spectral statistics that we extract are independent of $\omega_0$.
However, given the lack of such symmetry in the impurity model, in principle one has to consider its whole spectrum.
In practice, given the infinitely large bosonic Hilbert space of the cavity, we truncate it to a finite number of excitations $n_{\rm cutoff}=2^{10}$. We use standard algorithms with double precision.
To provide statistics that are converged with respect to $n_{\rm cutoff}$, we discard the upper $50 \%$ of the impurity eigenvalues.
Additionally, contrary to the lattice model whose spectrum was found to be statistically uniform throughout, the spectrum of the impurity model can be mixed: a low-energy portion with integrable statistics, and an intermediate to high-energy portion with chaotic statistics. Such features where already reported for similar models~\cite{emary2003chaos2,magnani2014comparative2,carlos2019quantum,das2022revisiting} and are consistent with the classical analysis presented in~\ref{classical}.
Hence, we focus on an intermediate energy range, discarding about the first $10 \%$ of the spectrum.

\subsection{Level-spacing statistics}
\begin{figure}[!tbp]
\centering
\includegraphics[width=4.0cm]{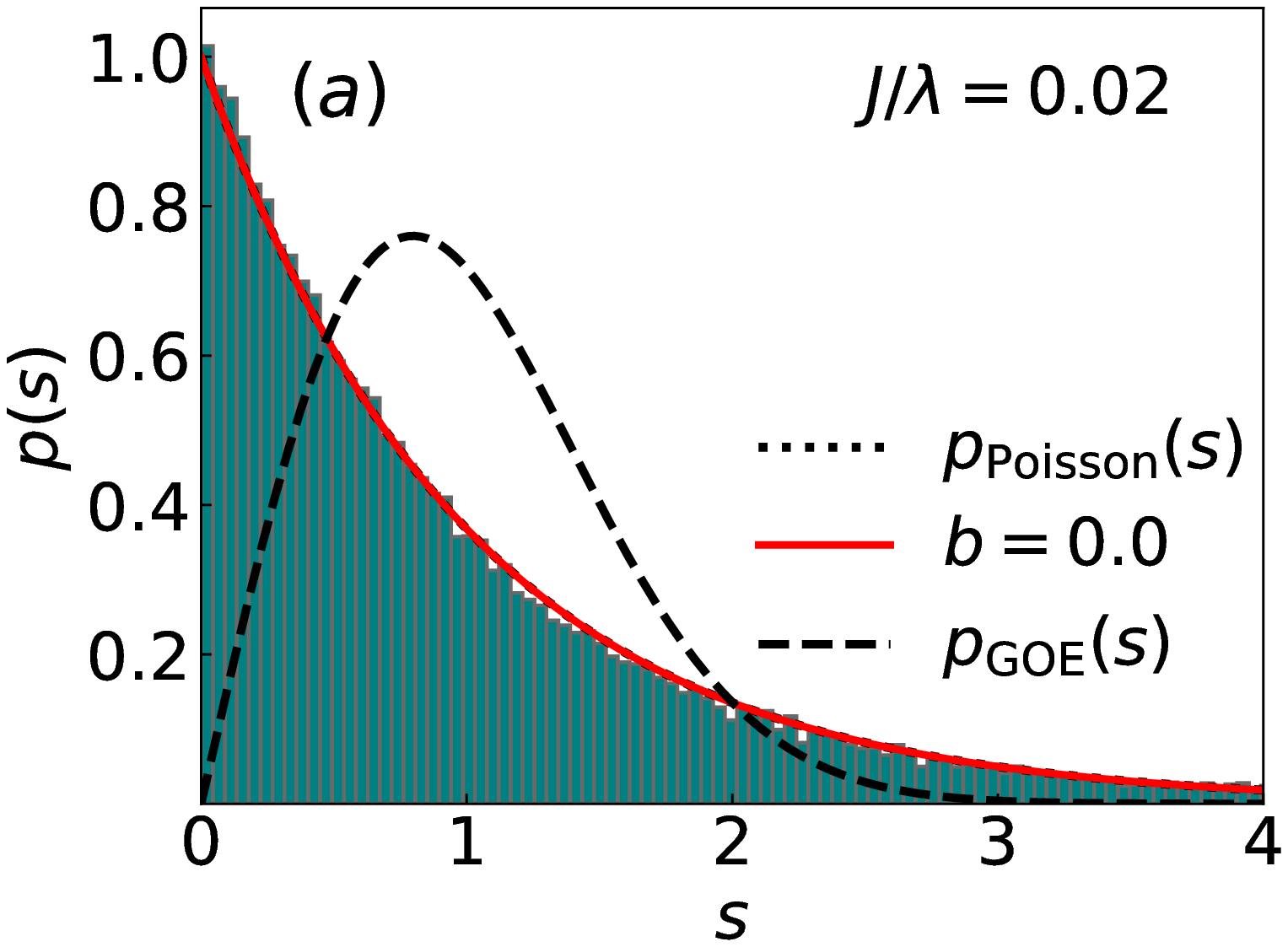}
\hspace{-1em}
\includegraphics[width=4.0cm]{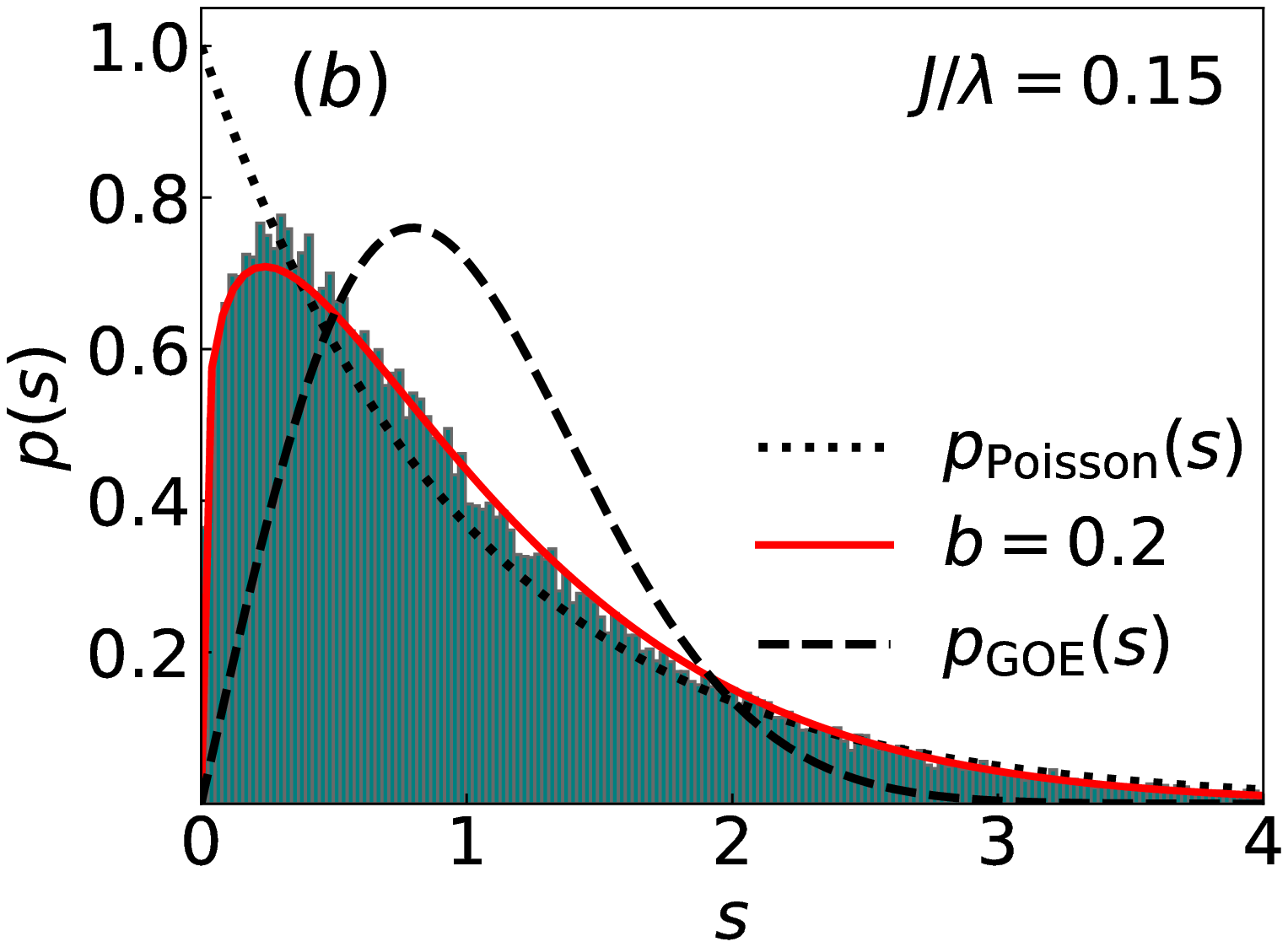}
\hspace{-1em}
\includegraphics[width=4.0cm]{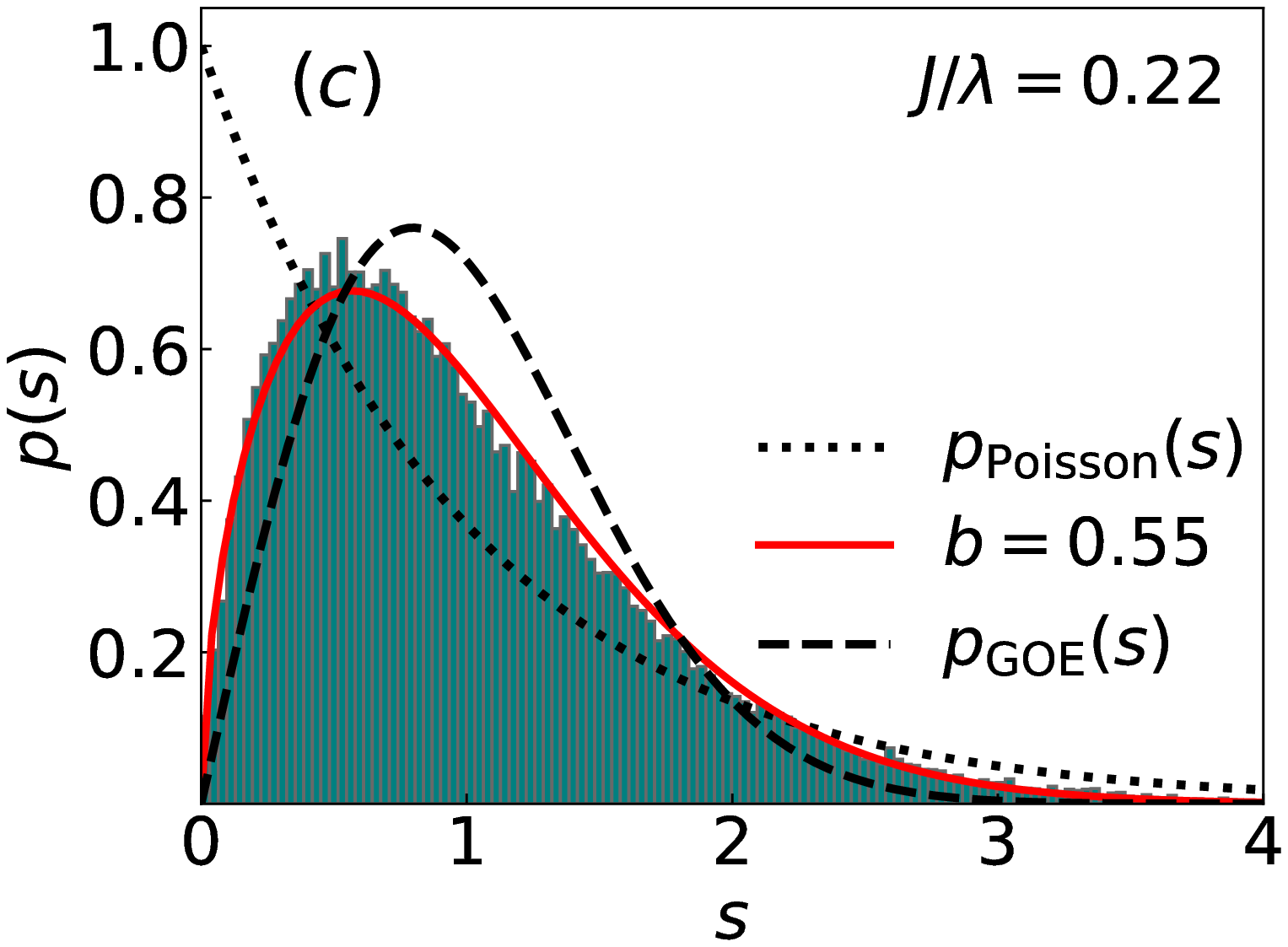}
\hspace{-1em}
\includegraphics[width=4.0cm]{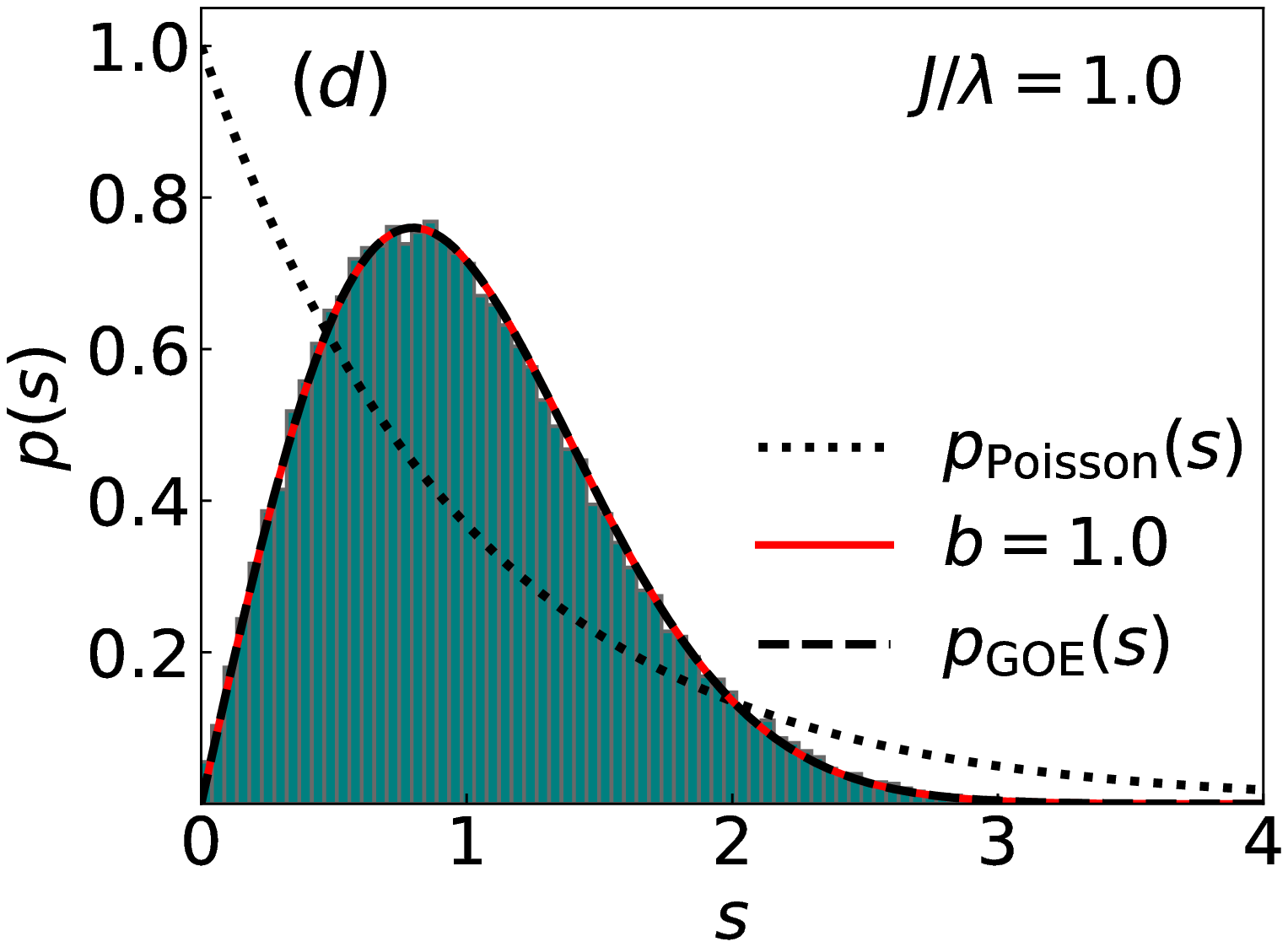}
\hspace{-1em}
\includegraphics[width=4.0cm]{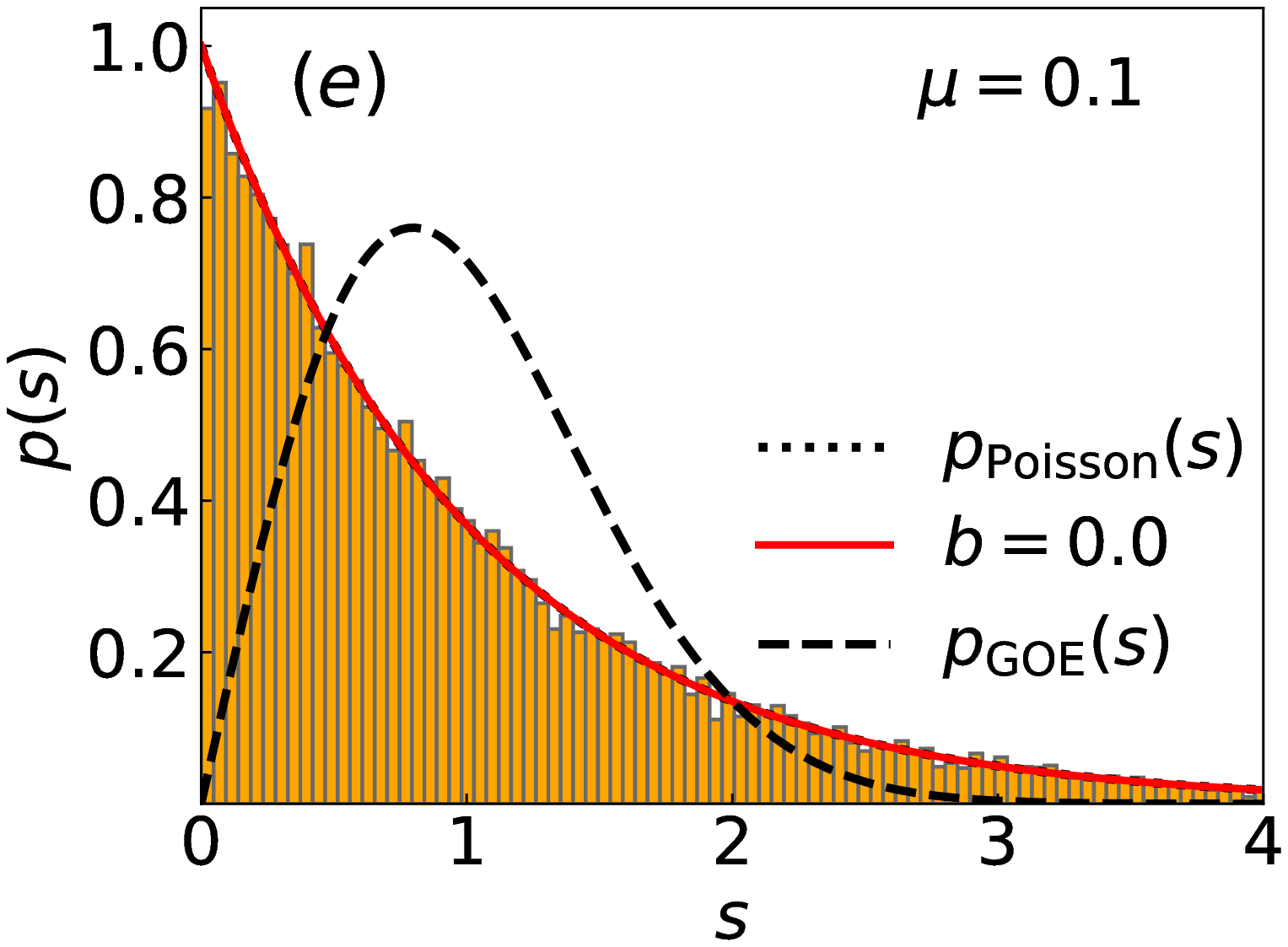}
\hspace{-1em}
\includegraphics[width=4.0cm]{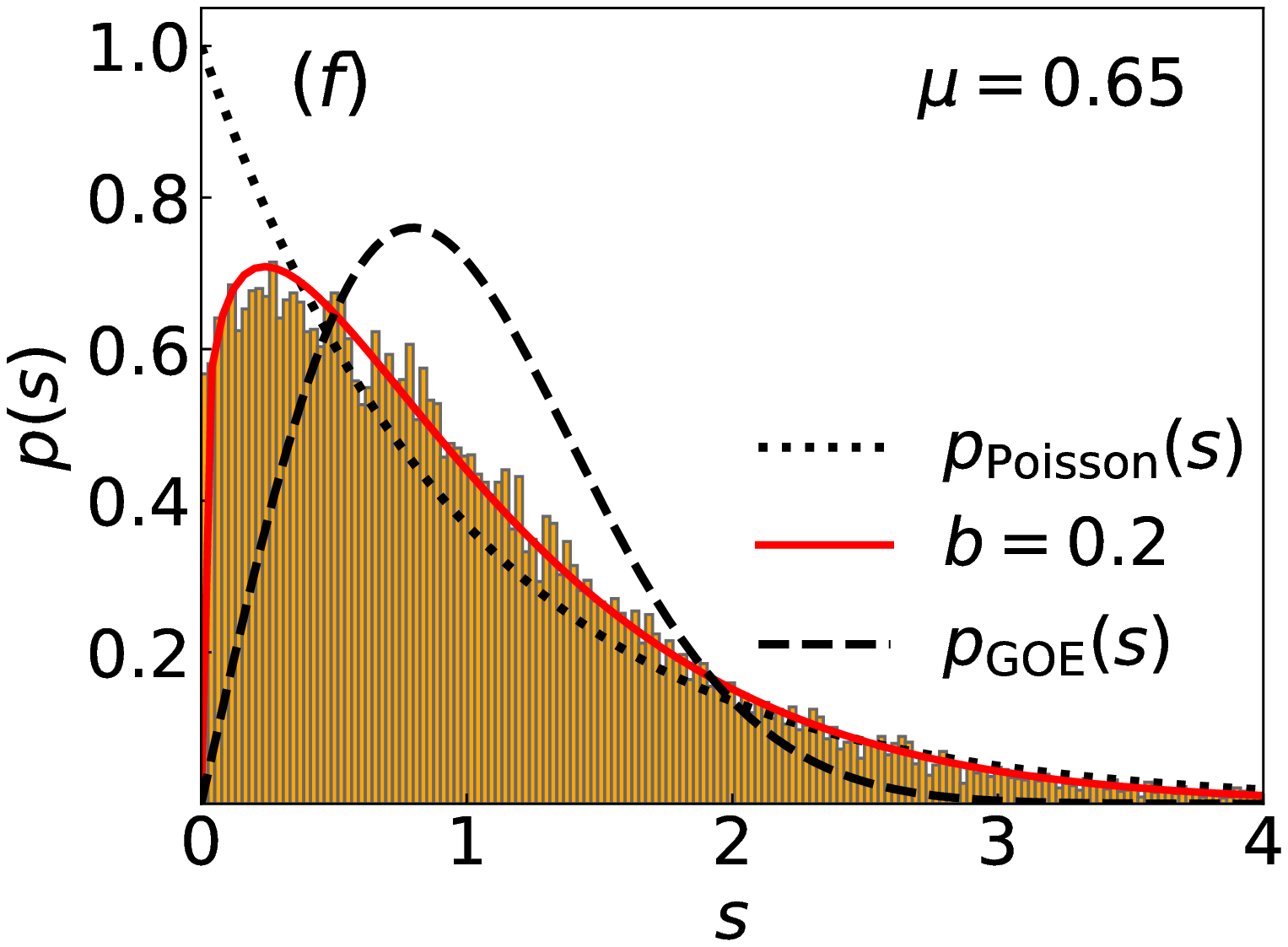}
\hspace{-1em}
\includegraphics[width=4.0cm]{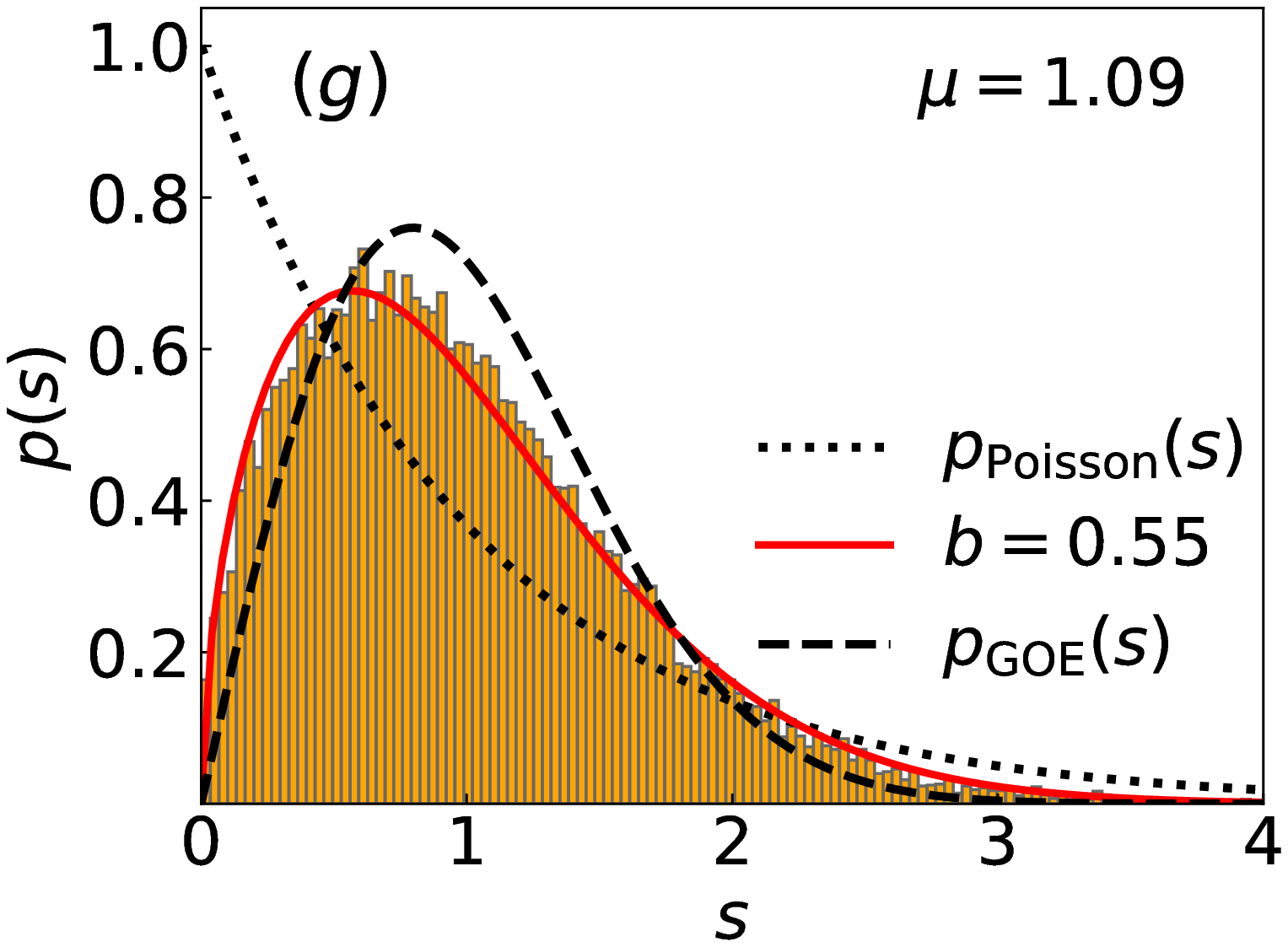}
\hspace{-1em}
\includegraphics[width=4.0cm]{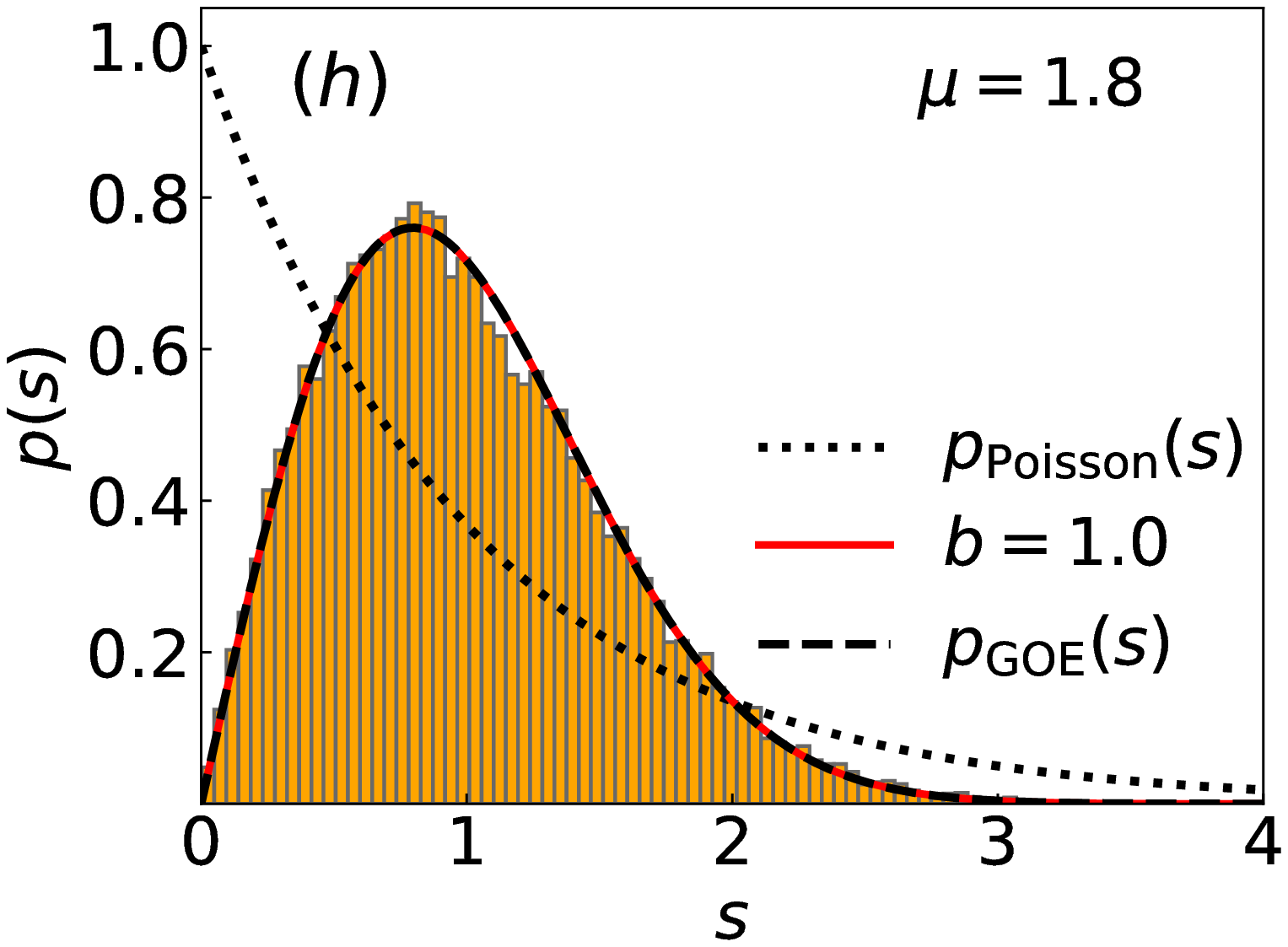}
\caption{Distribution of level spacings {$p(s)$ in the crossover regime from integrability to chaos (left to right)}.
The top panel is computed from the exact diagonalization of the Tavis-Cummings lattice ($L=3$ sites) with spin $S=8$, for
(a) $J/\lambda =0.02$, (b)  $J/\lambda =0.15$, (c) $J/\lambda =0.22$ and (d) $J/\lambda = 1.0$.
The bottom panel is computed from the corresponding impurity model with $S=64$ and $\lambda = 1$, for
(e) $\mu = 0.1$, (f) $\mu = 0.65$, (g) $\mu = 1.09$ and
(h) $\mu = 1.8$. 
The red curves correspond to fits to the Brody distribution defined in Eq.~(\ref{eq:brody}), with the single fitting parameter $b$ given in the legend.
The values of the impurity $\mu$ are chosen such that $b$ is the same between the lattice and the impurity.
}
\label{fig:level_spacing0}
\end{figure}

In order to unveil the universal footprints of these spectra {as well as the crossover regime between integrability and chaos}, we study the level-spacing statistics {of the TCL and its associated impurity model as their respective integrability-breaking parameters are turned on}.
First, we perform an unfolding of the spectra using standard procedures, see the details in~\ref{unfold}. 
The unfolded spectra are then used to generate the histograms of the gaps $s$ between nearest-neighbor eigenvalues, yielding the spacing distributions $p(s)$.
The results obtained for the lattice model are summarized in the top panel of Fig.~\ref{fig:level_spacing0} for weak, intermediate, and strong values of the integrability-breaking parameter $J/\lambda$.
For comparison, we also plot the corresponding spacing distributions for independent random numbers, namely the Poisson distribution~\cite{berry1977level}
\begin{align}
\label{eq:2dp}
p_{\rm Poisson}(s)=  \exp(-s) \,,
\end{align}
as well as the corresponding distribution for the eigenvalues of Hermitian random matrix ensemble~\cite{bohigas1984characterization}, namely the Gaussian Orthogonal Ensemble (GOE), 
\begin{align}
\label{eq:ginue}
 p_{\rm GOE}(s) = \frac{\pi}{2} s \, \exp\left(-\frac{\pi}{4} s^2\right)  \,.
\end{align}
The distributions computed from the spectrum of $H$ are in remarkable agreement with Poisson in the weak hopping regime, and with the GOE RMT prediction in the strong hopping regime.
The case of the Jaynes-Cummings lattice ($S=1$) has been studied at small filling fraction in Ref.~\cite{li2021eigenstate}.

We also compute the level-spacing statistics of the impurity model in both the integrable and RMT regime, as well as in the intermediate crossover regime. See the bottom panel of Fig.~\ref{fig:level_spacing0}. The impurity statistics succesfully reproduce the ones found on the lattice side in all these regimes.
{
Let us better quantify the agreement between the TCL and its associated impurity model by performing a single-parameter numerical fit of all the computed spacing distributions to the following Brody distribution
\begin{equation}
		\label{eq:brody}
P_B (b,s)=(b+1)\eta s^b \rme^{-\eta s^{b+1}}, \ \ \ \eta=\Gamma\left(\frac{b+2}{b+1}\right)^{b+1},
\end{equation}  
where $\Gamma(x)$ is the gamma function. 
The transition from RMT to Poisson statistics has already been extensively studied in a variety of models~\cite{TProsen_1993,kota2001random,schweiner2017,shukla2018,GARCIAGARCIA2004,Kravtsov_2015}
and the Brody distribution was heuristically proposed to interpolate between the Poisson and the GOE regimes~\cite{brody1981,TProsen_1993}.
We found it to be better suited than other interpolating distributions such as the one corresponding to the Rosenzweig-Porter for $2 \times 2$ matrices.
The results of the fitting procedure are given by the red curves in Fig.~\ref{fig:level_spacing0} and the corresponding values of $b$ are given in the legends.
The values of $\mu$ were chosen so as to reproduce the same values of $b$ as that of the lattice.

\subsection{Adjacent-gap ratio}

\begin{figure}[!tbp]
\centering
\includegraphics[scale=0.5]{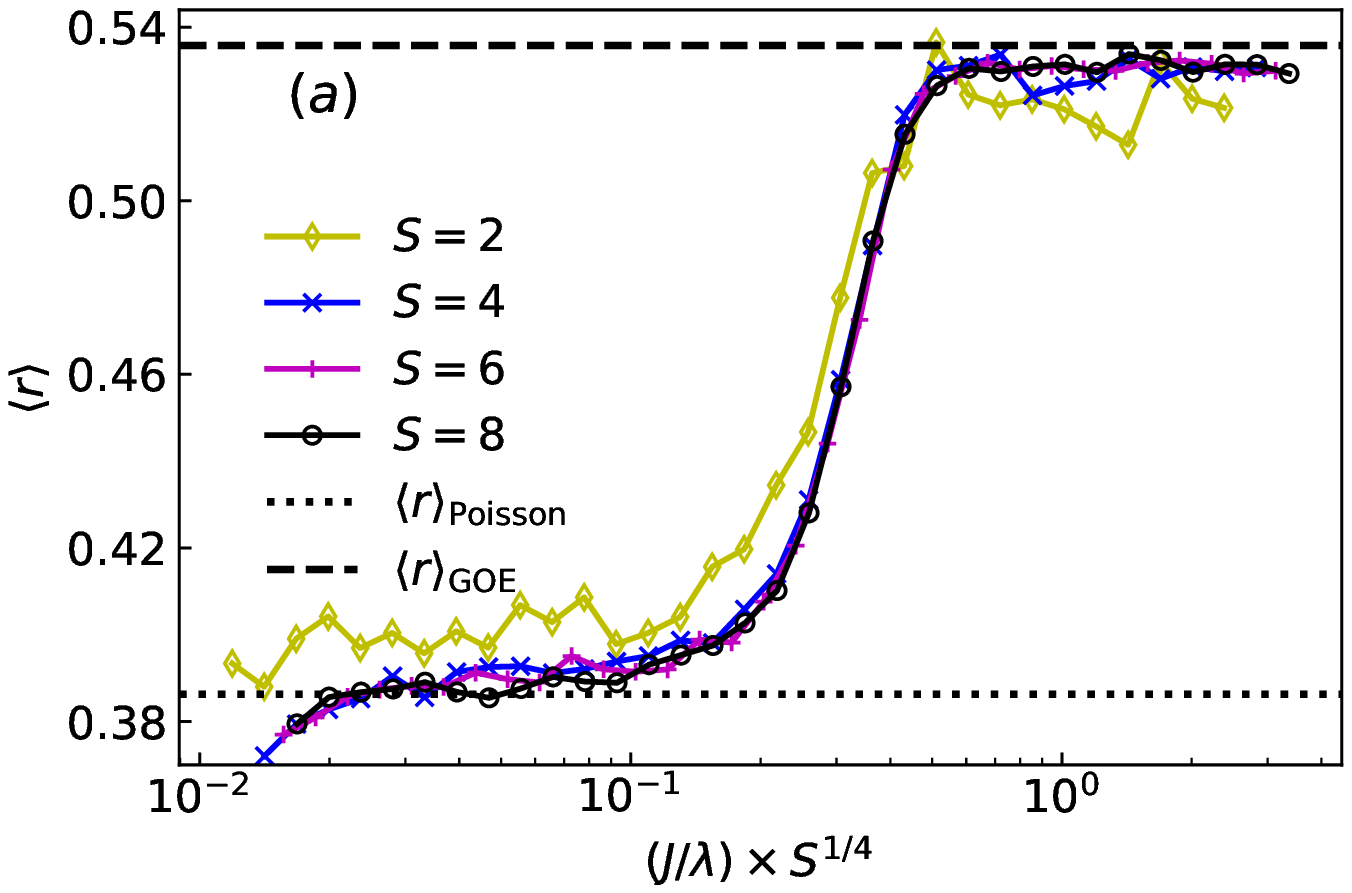}
\includegraphics[scale=0.5]{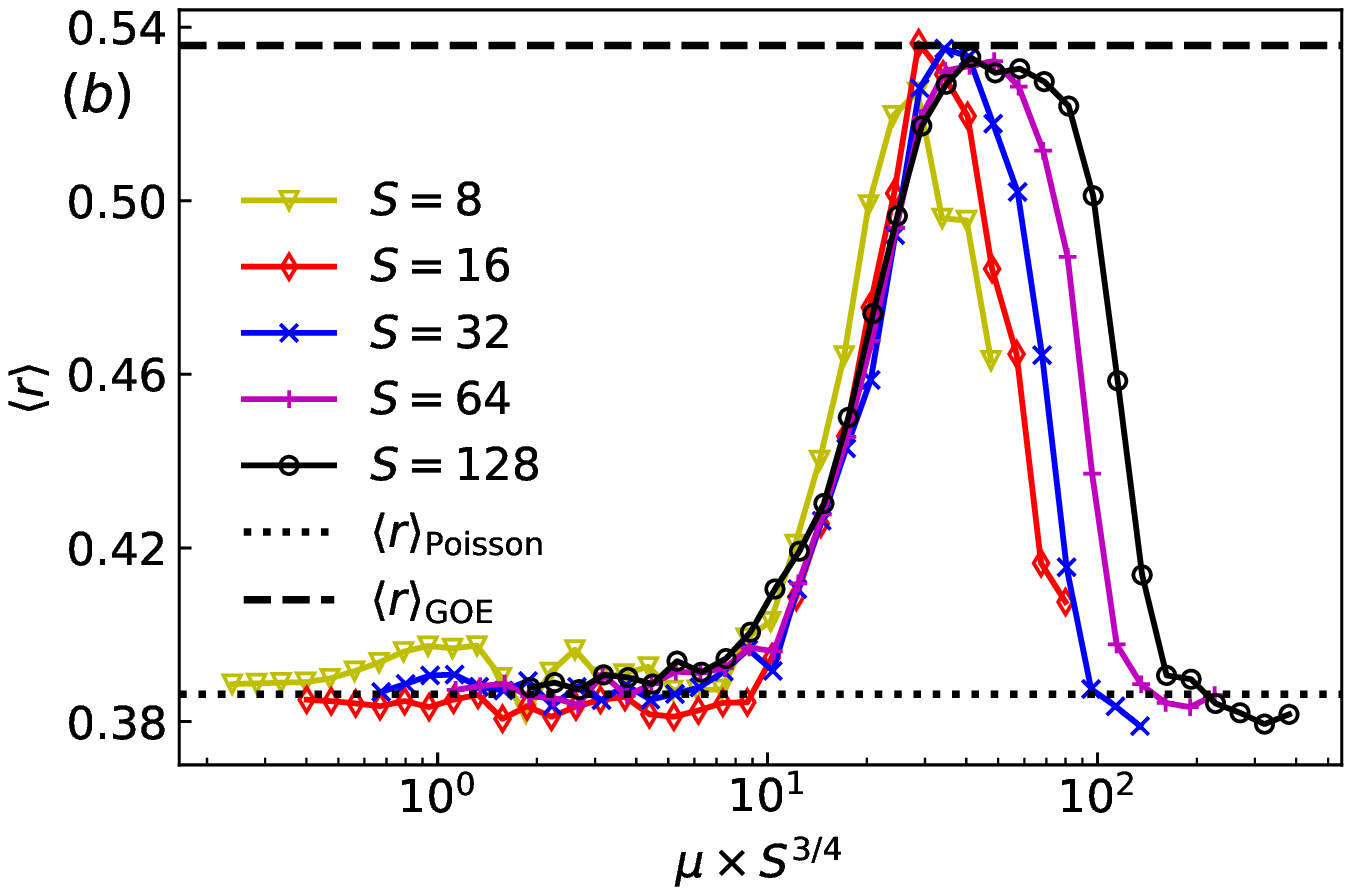}
\caption{
Adjacent gap ratio $\langle r \rangle$ {in the crossover regime from integrability to chaos}.
(a) Tavis-Cummings lattice ($L=3$ sites) as {$J/\lambda \times S^{1/4}$} is tuned from weak to strong hopping.
(b) Corresponding impurity model as a function of the drive {$\mu \times S^{3/4}$} for fixed $\lambda =1$. Other choices of $\lambda$ yield similar results. {The collapse of the different curves is used to identify the dynamic scaling with respect to local spin size $S$.}
\label{fig:gap_ratio}
}
\end{figure}
{As a complementary diagnostic to the spectral} statistics, we compute the adjacent-gap ratio~\cite{oganesyan2007localization} which does not rely on the unfolding procedure. It is defined as
\begin{align}
r_n = \frac{  \min (\delta_n, \, \delta_{n+1}) } {\max (\delta_n, \, \delta_{n+1}) }\,, \label{eq:gap_ratio}
\end{align}
where $\delta_n = E_{n+1} - E_n$ is the level spacing between two consecutive eigenvalues. 
For chaotic systems in the GOE class, the tabulated average adjacent gap ratio is $\langle r \rangle_{\rm GOE} \approx 0.53$.
For integrable cases,  $\langle r \rangle_{\rm Poisson}  \approx 0.39$.
In Fig.~\ref{fig:gap_ratio}~(a), we report how $\langle r \rangle$ evolves as a function of {the integrability-breaking parameter} $J/\lambda$ {and we identify its scaling with $S$. 
The figure displays a well delineated ramp where $\langle r \rangle$ crosses over from Poisson value to GOE value. We find a good collapse of that crossover ramp for different values of $S$ when $\langle r \rangle$ is plotted as a function of $J/\lambda \times S^{1/4}$. Interestingly, this implies that different sets of model parameters  will produce the same adjacent gap ratio as long as $J/\lambda \times S^{1/4}$ is kept constant. More generally, this scaling has to be interpreted in the sense of the dynamic scaling introduced in Ref.~\cite{bulchandani2021onset}: it is the scaling which allows to control the onset of chaos when the size of the local Hilbert space is increased.
}
Before the crossover ramp, the integrable phase is found to be robust until a finite value of  $J/\lambda {\times S^{1/4}} \approx 0.1$.
{After the crossover ramp, we observe a large chaotic plateau where the value of $\langle r \rangle$ is the one  of the GOE ensemble.}
{For values of {$J/\lambda \times S^{1/4}$} that are much larger than the ones shown in Fig.~\ref{fig:gap_ratio}~(a), {we found a departure from the RMT statistics. This is expected when} the kinetic energy dominates over the Tavis-Cummings light-matter coupling and the model perturbatively reduces to a free bosonic tight-biding model which is integrable}.

In Fig.~\ref{fig:gap_ratio}~(b), we display the same quantity on the impurity side as a function of {the integrability-breaking parameter $\mu$ and we identify its scaling with $S$. The impurity model successfully reproduces the qualitative features found in the lattice model: a robust integrable regime, a crossover ramp, and a subsequent chaotic plateau. We find a good scaling collapse of that ramp when $\langle r \rangle$ is plotted as a function of $\mu \times S^{3/4}$. As it will become clear later when relating both models, such a dynamics scaling of the impurity model is consistent with the one found in the lattice model.
At large values of {$\mu \times S^{3/4}$}, the re-entrance of the integrable phase 
can be attributed to the effective screening of the interaction in the Hamiltonian~(\ref{eq:H_imp}) by a very strong drive term and is consistent with what we observed on the lattice side at very large $J/\lambda \times S^{1/4}$.}
For both lattice and impurity models, we also checked that the entire adjacent-gap ratio distribution $P(r)$~\cite{atas2013distribution,atas2013joint,
giraud2022probing} converges to those universal distributions expected in the integrable and the chaotic regimes.

}

\subsection{Map between lattice and impurity models} \label{sec:map}
\begin{figure}
	\centering
	\includegraphics[scale=0.5]{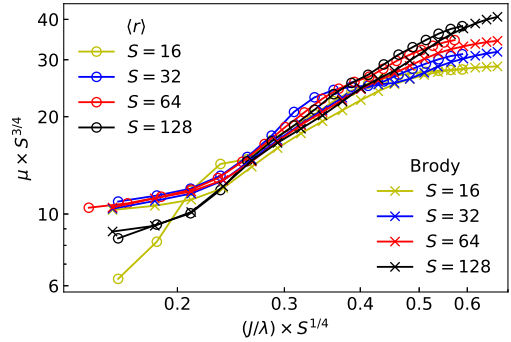}
\caption{
{Map between the integrability-breaking parameters of the lattice and impurity models, $J/\lambda$ and $\mu$, respectively. The scalings with $S$ on both axis are deduced from the dynamic scalings identified in Fig.~\ref{fig:gap_ratio}. The map is numerically extracted from the level-spacing Brody parameters $b$ determined in Fig.~\ref{fig:level_spacing0} as well as from the data of the adjacent gap ratio $\langle r \rangle$ of Fig.~\ref{fig:gap_ratio}. 
The methodology is explained in Sec.~\ref{sec:map}.
$L=3$ and $S=8$ on the lattice side and the different impurity spin sizes are given in the legend.}
	\label{fig:mu_J}}
\end{figure}
{
Above, we have characterized the integrable-to-chaotic crossover of the level-spacing distribution via its Brody parameter $b$, as well as the one of the adjacent gap ratio $\langle r \rangle$ when increasing the integrability-breaking parameters, namely $J/\lambda$ in the lattice and $\mu$ in the impurity. We use this to retrieve the relationship between the integrability-breaking parameters of both models by eliminating the fitting parameter $b$ from the data of Fig.~\ref{fig:level_spacing0} and, similarly, by eliminating $\langle r \rangle$ from the data of Fig.~\ref{fig:gap_ratio}.
In practice, we identify the functions $\alpha=f_1(J/\lambda)$ and $\alpha=f_2 (\mu)$ where $\alpha = b, \langle r \rangle$ and find $\mu=f_2^{-1} (f_1 (J/\lambda))$. Note that these functions are only invertible in the crossover region.
The resulting maps $\mu(J/\lambda)$ for different impurity sizes $S$ are displayed in Fig.~\ref{fig:mu_J} for a fixed lattice size $L=3$ and $S=8$. We use the dynamic scaling  identified in Fig.~\ref{fig:gap_ratio}. The resulting maps obtained from the level-spacing distributions are similar to those obtained from the adjacent gap ratio, and the agreement seems to improve as $S$ is increased.
Notably, this reveals that the impurity $\mu$ is a non-linear function of the lattice $J/\lambda$.
From a mean-field point of view, this map has to be understood as the self-consistent relation tying together the TCL model with its impurity counterpart.
}

\subsection{Spectral Form Factor}

\begin{figure}[!tbp]
	\centering
	\includegraphics[scale=0.25]{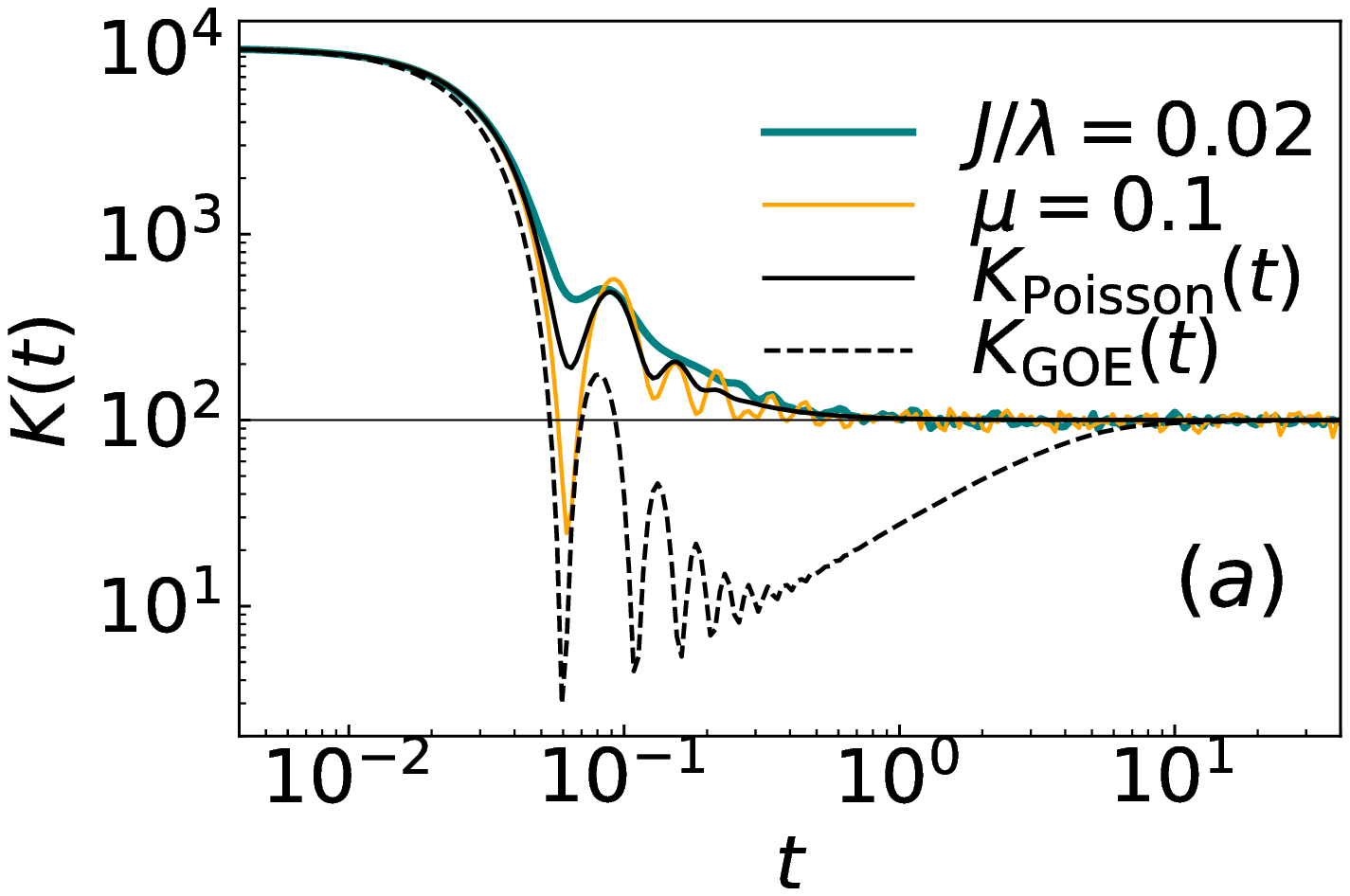}
\hspace{-0.7em}
	\includegraphics[scale=0.25]{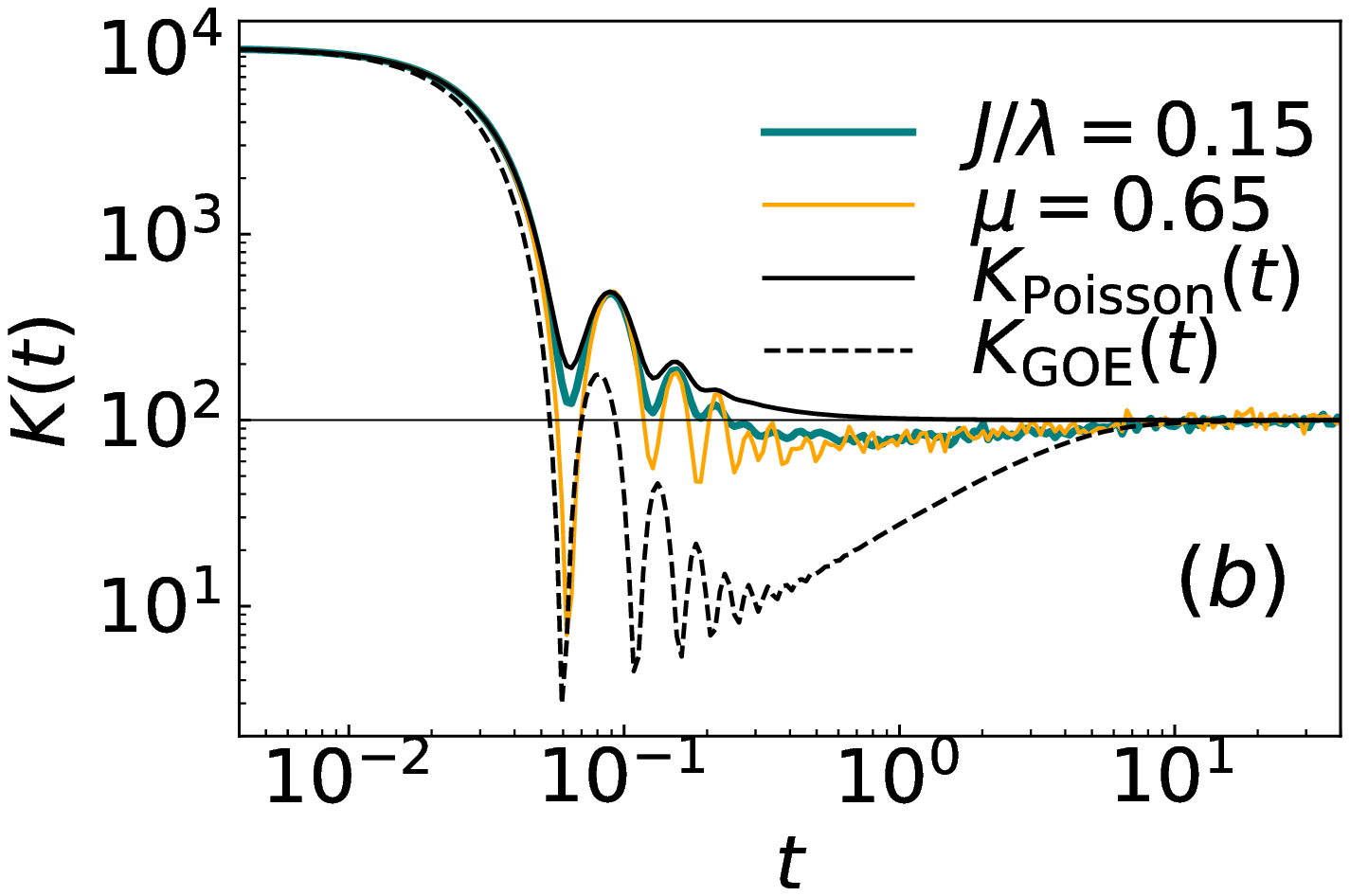}
\hspace{-0.7em}	
	\includegraphics[scale=0.25]{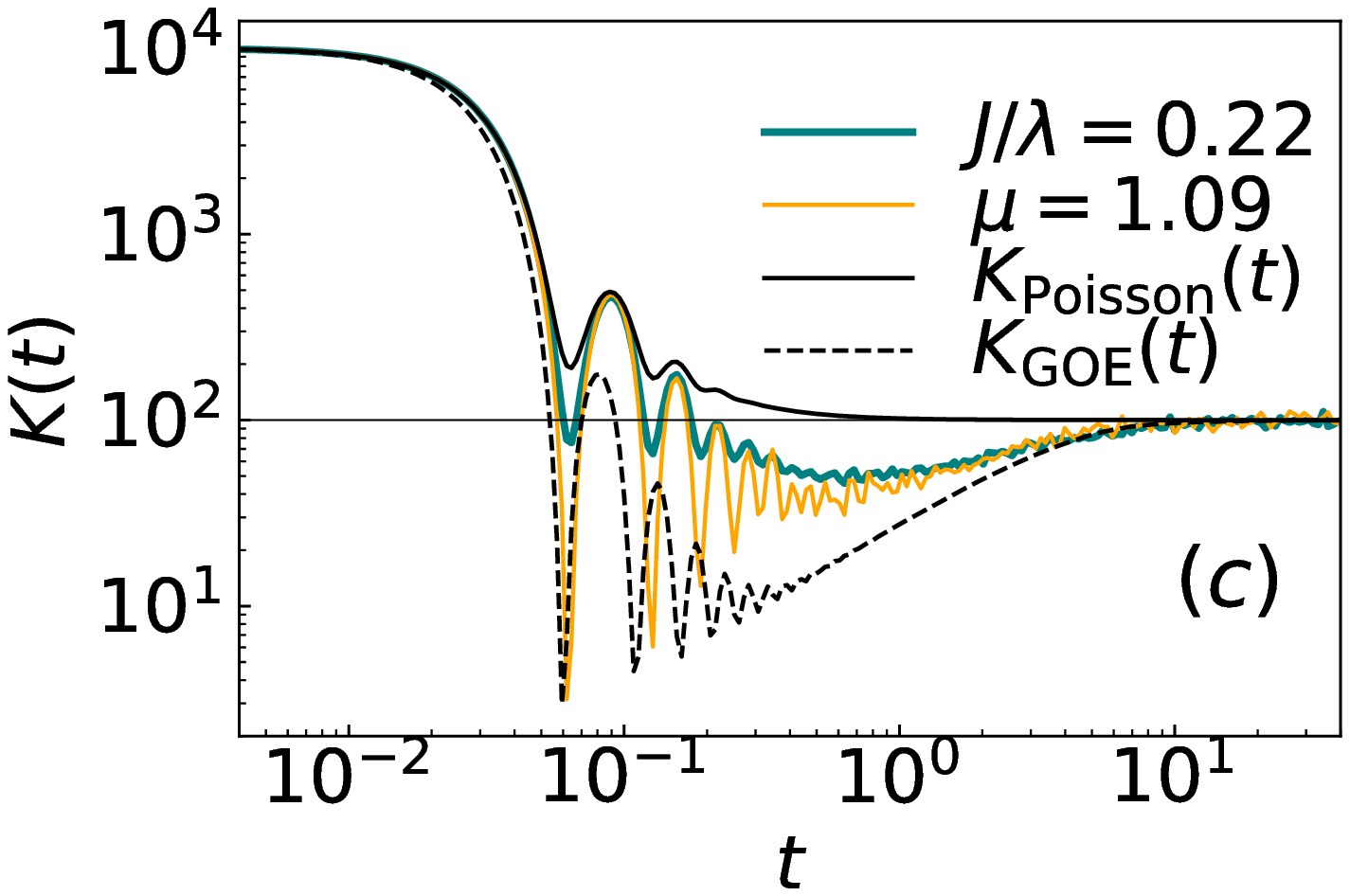}
\hspace{-0.7em}
	\includegraphics[scale=0.25]{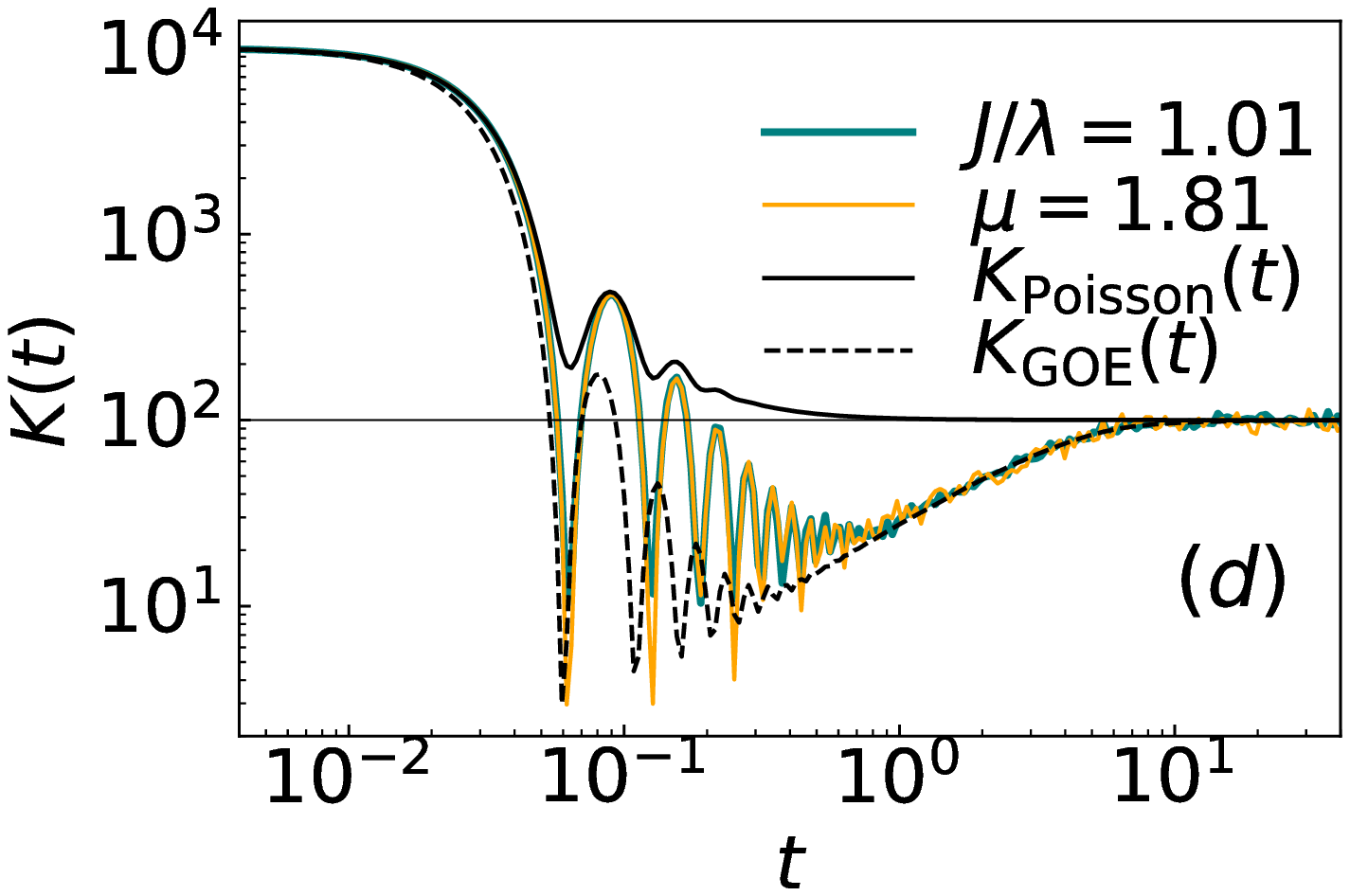}
	\caption{
Spectral form factors of the lattice model with $L=3$ and $S=8$ and of the impurity model with $S=64$ for increasing values of the integrability-breaking parameters $J/\lambda$ (lattice) and $\mu$ (impurity) given in the legends and chosen as in Fig.~\ref{fig:level_spacing0}.
The plain and dashed black lines correspond to the Poisson and GOE distributions, respectively.
		\label{fig:SFF}
	}
\end{figure}
We now turn to another diagnostic which probes the long-range correlations in the spectrum, namely the spectral form factor (SFF) that is defined as~\cite{haake2010quantum}
\begin{align}
K(t)  = \big{\langle} \left| \mathrm{Tr} \, \rme^{\rmi H t} \right|^2 \big{\rangle} =  \big{\langle} \sum_{m, \, n=1}^N \rme^{\rmi(E_m - E_n) t} \big{\rangle} \,.
\end{align}
 $\langle \ldots \rangle$ typically denotes averaging with respect to disorder sampling.
For our clean system, we replace this disorder average by dividing the unfolded spectrum into samples of $N =100$ consecutive eigenvalues and by averaging over those samples.
This is justified by the expectation that the statistics are similar throughout the spectrum.
The resulting SFF for the lattice and the impurity model are presented in Fig.~\ref{fig:SFF} for the same parameters as in Fig.~\ref{fig:level_spacing0}.
We compare these findings to the predictions of the relevant random matrix ensembles~\cite{haake2010quantum,cotler2017chaos,gharibyan2018onset,
liu2018spectral,chen2018universal,bertini2018exact,kos2018many}. The SFF of the 
GOE RMT reads, {asymptotically at large $N$},
\begin{align}
K_{{\rm GOE}}(t)= & \left[\frac{\pi}{t}J_{1}(2Nt/\pi) \right]^2 \label{eq:Z_GOE}
 + N \begin{Bmatrix}
\frac{t}{\pi}-\frac{t}{2\pi} \log(1+\frac{t}{\pi}) \, , \ \ 0< t<2\pi 
\\2-\frac{t}{2\pi} \log(\frac{t+\pi}{t-\pi})  \, , \ \ 2\pi <t< \infty
\end{Bmatrix} . 
\end{align}
where $J_1(x)$ is the Bessel function of the first kind.
The early-time behavior of SFF, its dip and subsequent oscillations, are dominated by non-universal features of the spectrum.
In the intermediate to long-time regime, the linear ramp between Thouless and Heisenberg times and the subsequent plateau are well-known universal signatures of quantum chaos.
On the integrable side, the SFF for Poissonian levels with unit mean level 
spacing reads~\cite{prakash2021universal,riser2020nonperturbative}
\begin{align}
K_{{\rm Poisson}}(t)=N+\frac{2}{t^2}-\frac{(1+ \rmi t)^{1-N}+(1- \rmi t)^{1-N}}{t^2}\,. \label{eq:Z_P} 
\end{align}
In stark contrast to the chaotic case, the SFF of integrable dynamics does not show the linear ramp.
{In the fully developed integrable and chaotic regimes,} our results in Fig.~\ref{fig:SFF} are in excellent agreement with those universal predictions given in Eqs.~(\ref{eq:Z_GOE}) and~(\ref{eq:Z_P}). {More importantly,} this also clearly demonstrates that the SFF of the lattice and the impurity models are in quantitative agreement with each other {throughout the crossover region between the integrable and chaotic regimes}.

\section{Conclusion and Discussion}
In this work, we argued that the universal spectral features of a spatially extended system can be captured by a minimal impurity model with a much smaller Hilbert space.
This impurity modeling is inspired from what is routinely done to capture local physics in the \textit{thermodynamic} scaling regime.
Here, we proposed to extend this approach to the \textit{dynamic} scaling regime to capture spectral features {at the onset of chaos}.
The validity of this approach was tested by comparing spectral statistics computed on both the lattice and the impurity side. A complementary test would be to compare the chaotic features of out-of-time-order correlators.
Note that we have treated the integrability-breaking parameter $\mu$ as freely adjustable. 
This is similar to proving that a single spin coupled to a carefully chosen Weiss field is a faithful impurity representation of an extended magnet in that it can \emph{exactly} reproduce its magnetization.
However, an exciting challenge remains: \emph{analytically} identifying the relationship between the lattice problem and its impurity that self-consistently determines the amplitude of the integrability-breaking drive term. This means analytically deriving the map between $J/\lambda$ and $\mu$ that has been numerically computed in Fig.~(\ref{fig:mu_J}).
Although our elementary implementation relied on a single-site impurity driven by a static source, its generalization to larger impurities (\textit{e.g} to accommodate larger LIOMs, increase the local Hilbert space) or more complex environments is not expected to bring extra conceptual difficulty.
Adapting this approach to other lattice models relies on:
$(i)$ the impurity model featuring a tunable integrability-breaking term analogous to our $\mu$,
$(ii)$ a local (impurity) Hilbert space which is large enough to ensure sufficient  spectrum data for universal statistics to develop. 

As a side note, we found that both the quantum and the classical versions of our impurity model exhibit a rich phenomenology, with regimes of chaos and integrability simultaneously present at different energies, see the~\ref{classical}.
Similar observations were made in various other models~\cite{magnani2014comparative2,magnani2017regularity,lewis2019unifying,corps2022chaos,altland2012equilibration}. 
The classical-to-quantum correspondence of such models with mixed phase space is still an open question that could be investigated through the lens of an energy-resolved extension of BGS conjecture.

\section{Acknowledgments}
HKY, CA, MK are grateful for the support from the project 6004-1 of the Indo-French Centre for the Promotion of Advanced Research (IFCPAR). MK acknowledges the support of the Ramanujan Fellowship (SB/S2/RJN-114/2016), SERB Early Career Research Award (ECR/2018/002085) and SERB Matrics Grant (MTR/2019/001101) from the Science and Engineering Research Board (SERB), Department of Science and Technology, Government of India. MK acknowledges support of the Department of Atomic Energy, Government of India, under Project No. RTI4001. CA acknowledges the support from the French ANR ``MoMA'' project ANR-19-CE30-0020. MK thanks the hospitality of  \'Ecole Normale Sup\'erieure (Paris).

\appendix

\section{Unfolding the spectrum}
\label{unfold}
To eliminate the system-specific features of the spectrum, to extract its universal features, and to compare them with random matrix 
theory predictions, it is customary to perform a so-called unfolding procedure of the spectrum~\cite{french1971some, bohigas1991random,meyer1997random,bruus1997energy, guhr1999statistical,paar1991broken,guhr1998random,abul2014unfolding}. 
It proceeds by transforming the original spectrum such as to ensure a uniform local density of states in the resulting spectrum.
In practice, we use the following procedure:
\begin{itemize}
	\item[1.] First, we compute the cumulative density of the ordered spectrum, $I(E)= \sum _{n} \Theta (E-E_n)$ where $\Theta(x)$ is the Heaviside step function.
	\item[2.] $I(E)$ is then fitted to a smooth polynomial function $\tilde{I}(E)$.
	\item[3.] Finally, the unfolded spectrum is obtained as, $\tilde{E}_n \equiv \tilde{I}(E_n)$.
\end{itemize}

\begin{figure}[h]
	\centering
	\includegraphics[scale=0.3]{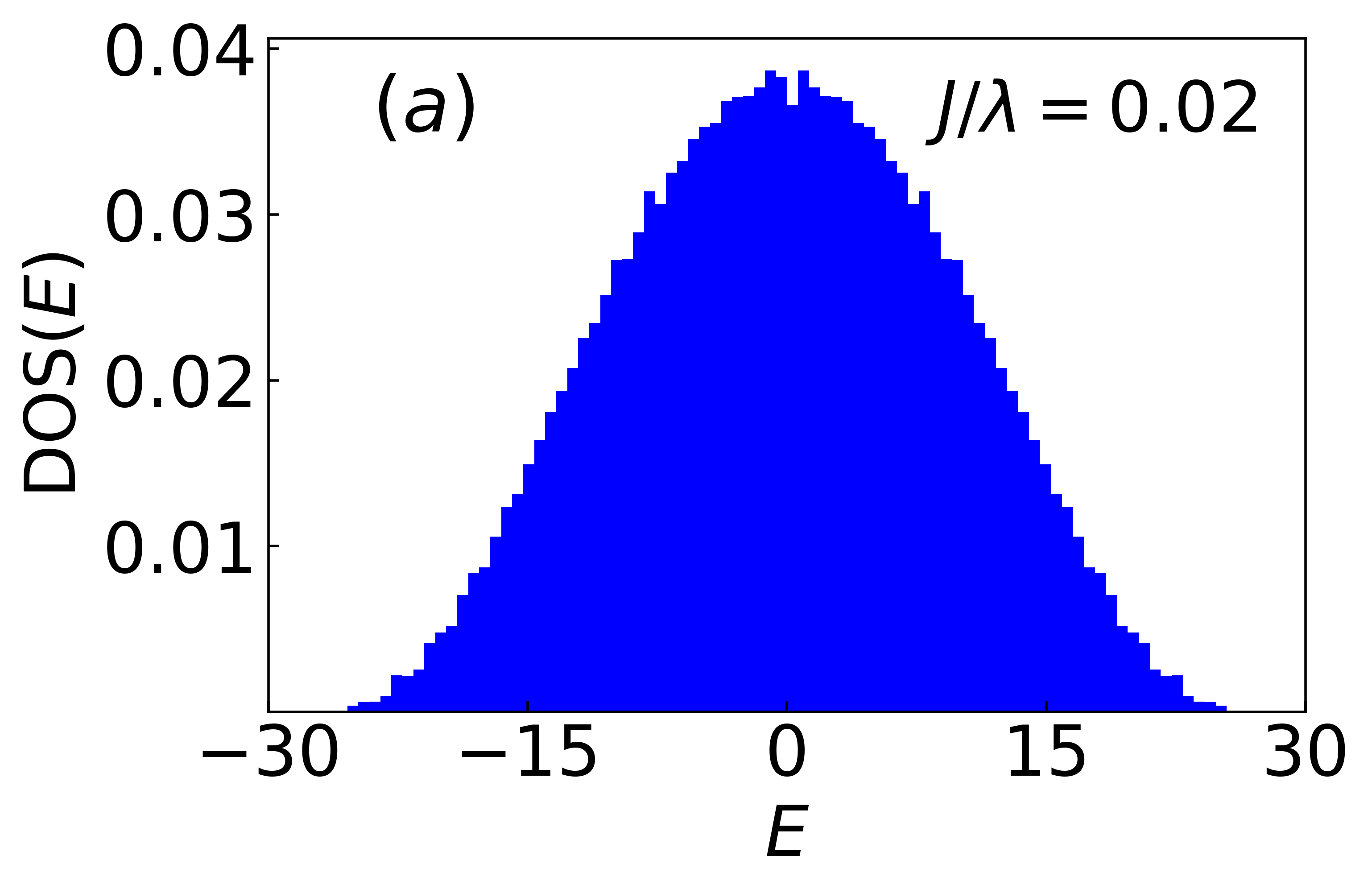}
	\includegraphics[scale=0.3]
	{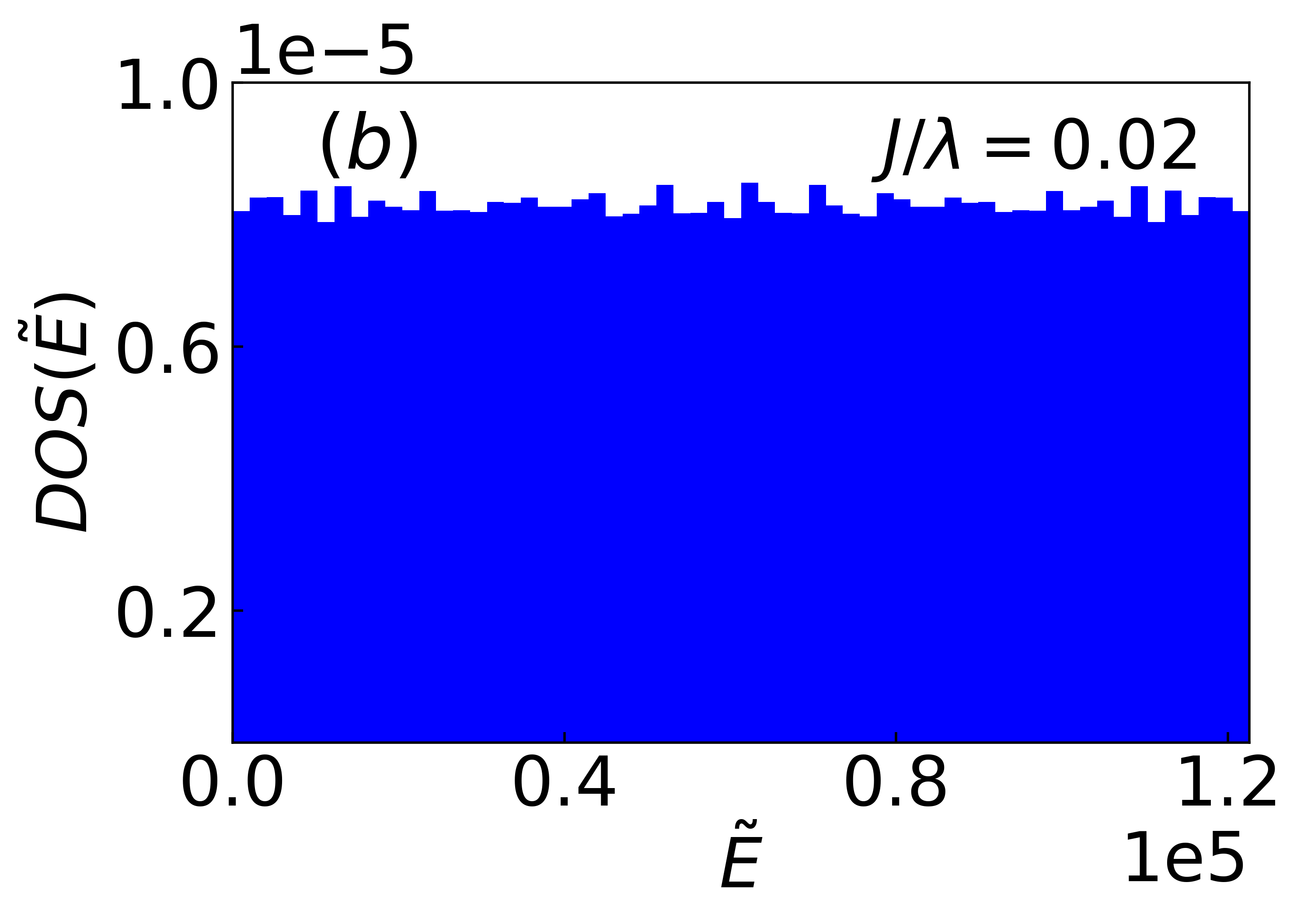}\\
	\includegraphics[scale=0.3]{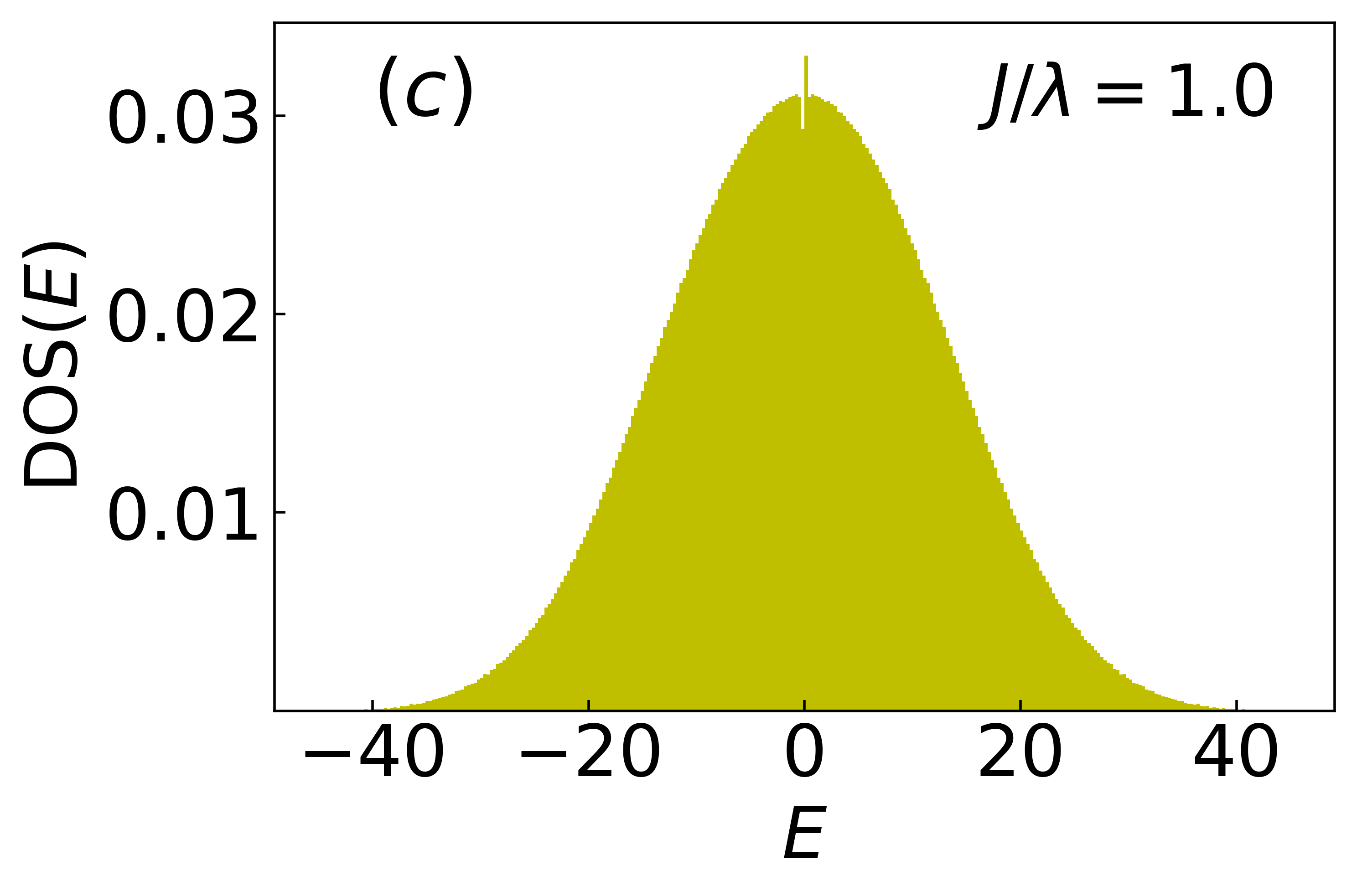}
	\includegraphics[scale=0.3]
	{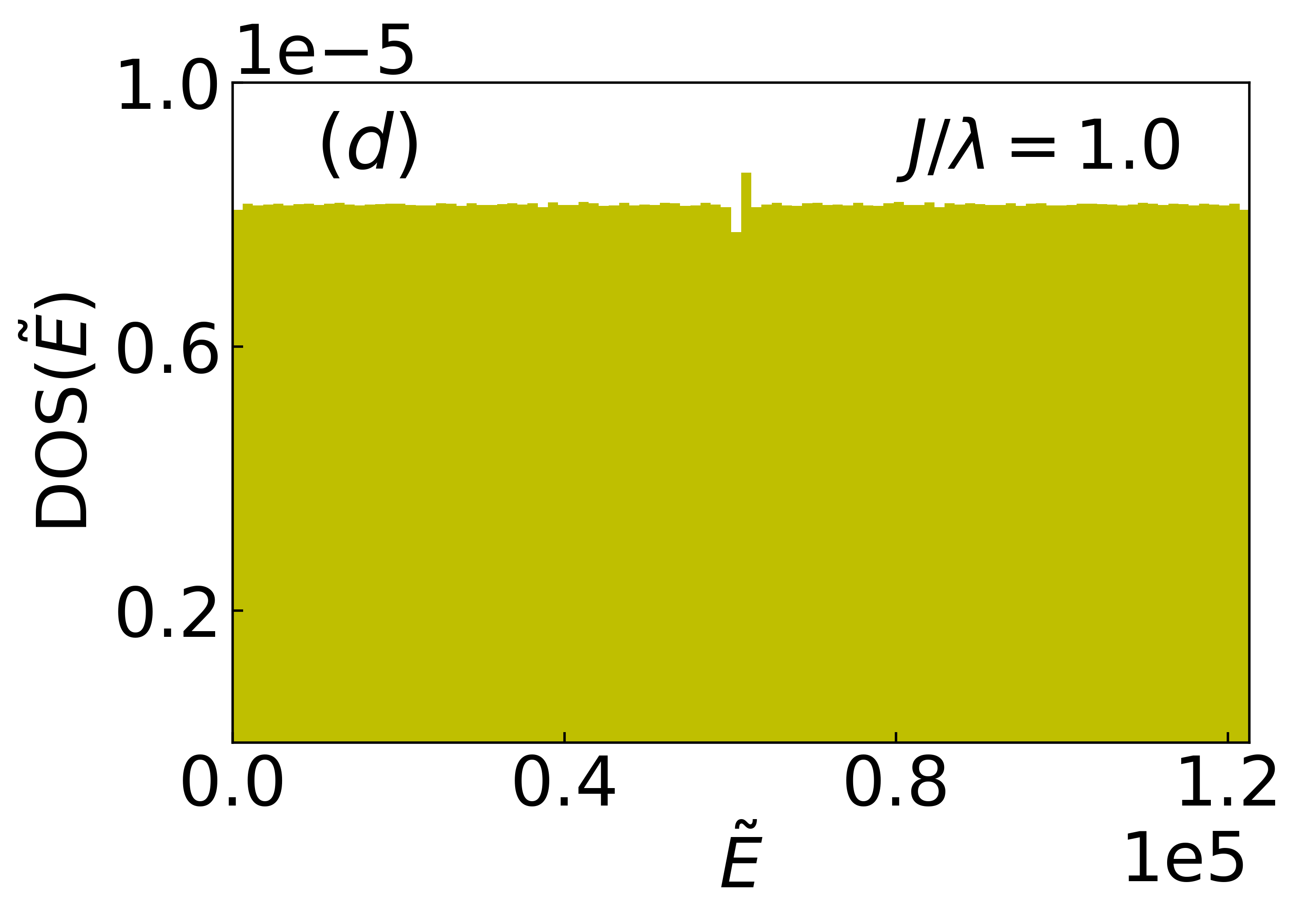}
	\caption{Density of states (DOS) of the Tavis-Cummings Lattice [Eqs.~(1-3) in the main text].
		(Left panel) Before unfolding.  (Right panel) After unfolding. Spectra are computed in the reflection-symmetric sector with $L=3,\ S=8$, $N_{\rm ex}=36$, and $\lambda=1.0$ for (a, b) weak hopping $J/\lambda=0.02$, and (c, d) strong hopping $J/\lambda=1.0$}
	\label{fig:DOS}
\end{figure}

We display in Fig.~(\ref{fig:DOS}) the density of states of the original spectrum and its corresponding unfolded spectrum for two different values of $J/\lambda$. We have used a 12$^{\rm th}$-order polynomial in the unfolding procedure. Clearly, the resulting density of states is almost constant.
The level-spacing distribution is computed from the unfolded spectrum as $p(s)=\sum_{n}^{}\delta(s-(\tilde{E}_{n+1}-\tilde{E}_{n}))$. After unfolding, the mean level spacing is unity by construction, $\langle s \rangle = \int_0^\infty {\rm d} s \, s\, p(s)  = 1$, and the higher moments are expected to display universal features depending on the integrable or chaotic nature of the dynamics.

\section{Adjacent gap ratio}
\label{agr}
The statistics of the unfolded spectrum may be sensitive to the precise procedure used to produce it.
To circumvent this shortcoming, one may resort to another statistical measure based on the ratios of adjacent gaps~\cite{oganesyan2007localization} which does not rely on an unfolding of the spectrum.
The distribution  of adjacent gap ratios for an ordered spectrum is defined as 
$P(r)=\sum_{n}\delta(r-r_n)$, where 
\begin{align}
	r_n = \frac{  \min (\delta_n, \, \delta_{n+1}) } {\max (\delta_n, \, \delta_{n+1}) }\,,
\end{align}
and $\delta_n = E_{n+1} - E_n$ is the level spacing between two consecutive eigenvalues. 
Clearly, $P(r)$ has support only in the interval $r \in \left[0,1\right]$.

Analytical expressions for the adjacent gap ratio distribution~\cite{atas2013distribution, atas2013joint, giraud2022probing} for integrable (independent Poisson numbers) and chaotic (RMT) spectra are given by 
\begin{equation}
	P_{{\rm Poisson/RMT}}(r)=2\,\tilde{P}_{{\rm Poisson/RMT}}(r)\Theta(1-r),
\end{equation}
with
\begin{align}
	\tilde{P}_{{\rm Poisson}}(r)=&\frac{1}{(1+r)^2} \,,\nonumber\\
	\tilde{P}_{{\rm RMT}} (r)=&\frac{1}{Z_\beta} 
	\frac{(r+r^2)^\beta}{(1+r+r^2)^{3/2+\beta}} \,,
	\label{eq:analytical_Pr}
\end{align}
where $Z_\beta$ is the normalization constant which depends on the Dyson index of the random 
matrix ensemble, $\beta$. For the Gaussian Orthogonal Ensemble (GOE),  $\beta=1$ and $Z_\beta=8/27$.

The average adjacent gap ratio, defined as $\langle r \rangle=\int_{0}^{1} {\rm d}r \, r P(r)$, is commonly used as a quantitative measure of quantum chaos and integrability.
It is especially useful for tracking the transition from chaos to integrability as a function of a Hamiltonian parameter.
The exact value of $\langle r \rangle$ can be computed using Eq. (\ref{eq:analytical_Pr}) for independent Poisson levels and GOE RMT:
\begin{align}
	\langle r \rangle_{\rm Poisson} = 2 \ln 2-1 \approx 0.386\,, \nonumber\\
	\langle r \rangle_{\rm GOE} = 4-2\sqrt{3} \approx 0.536\,.
\end{align}

\begin{figure}
	\centering
	\includegraphics[scale=0.3]{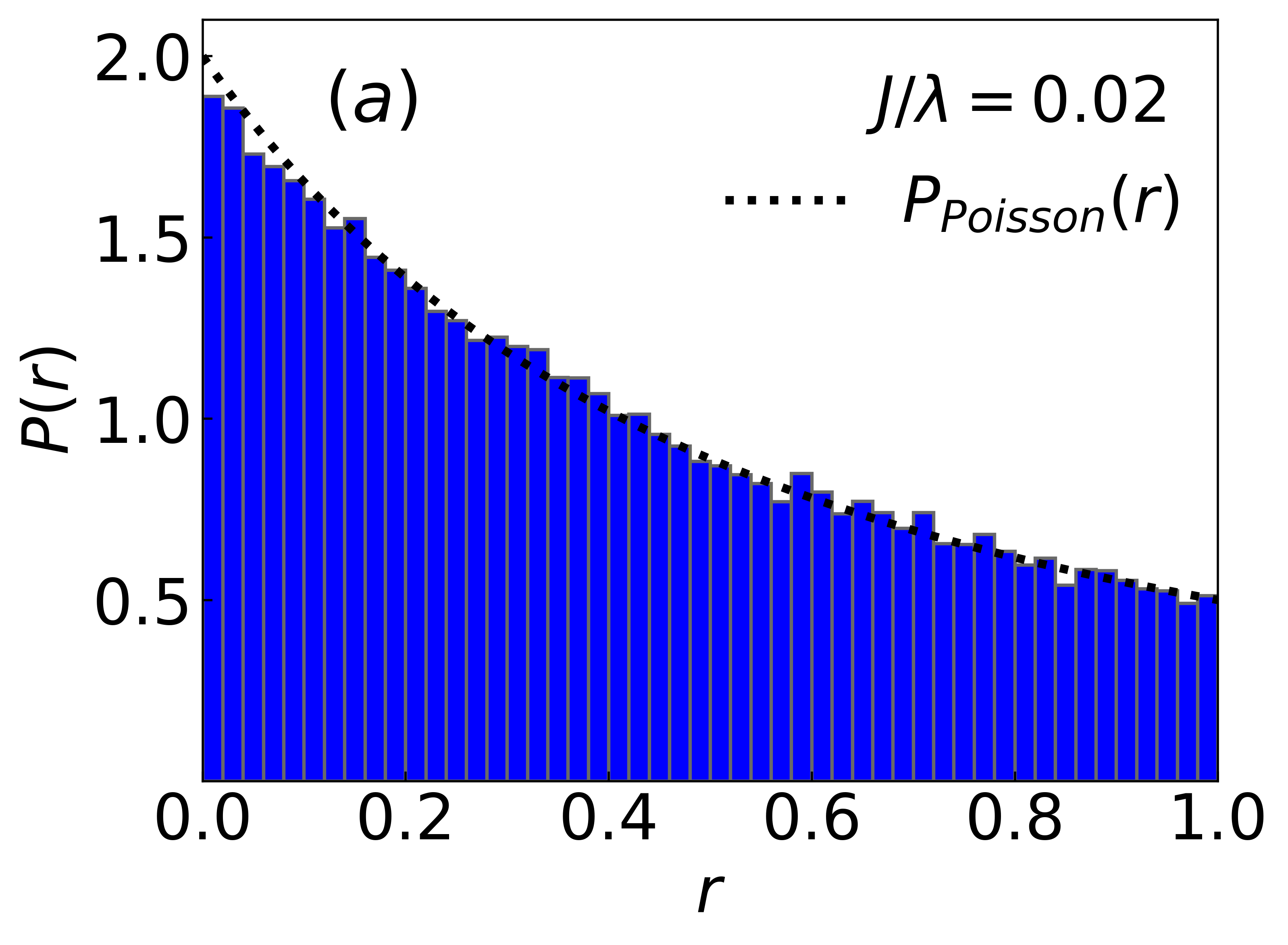}
	\includegraphics[scale=0.3]
	{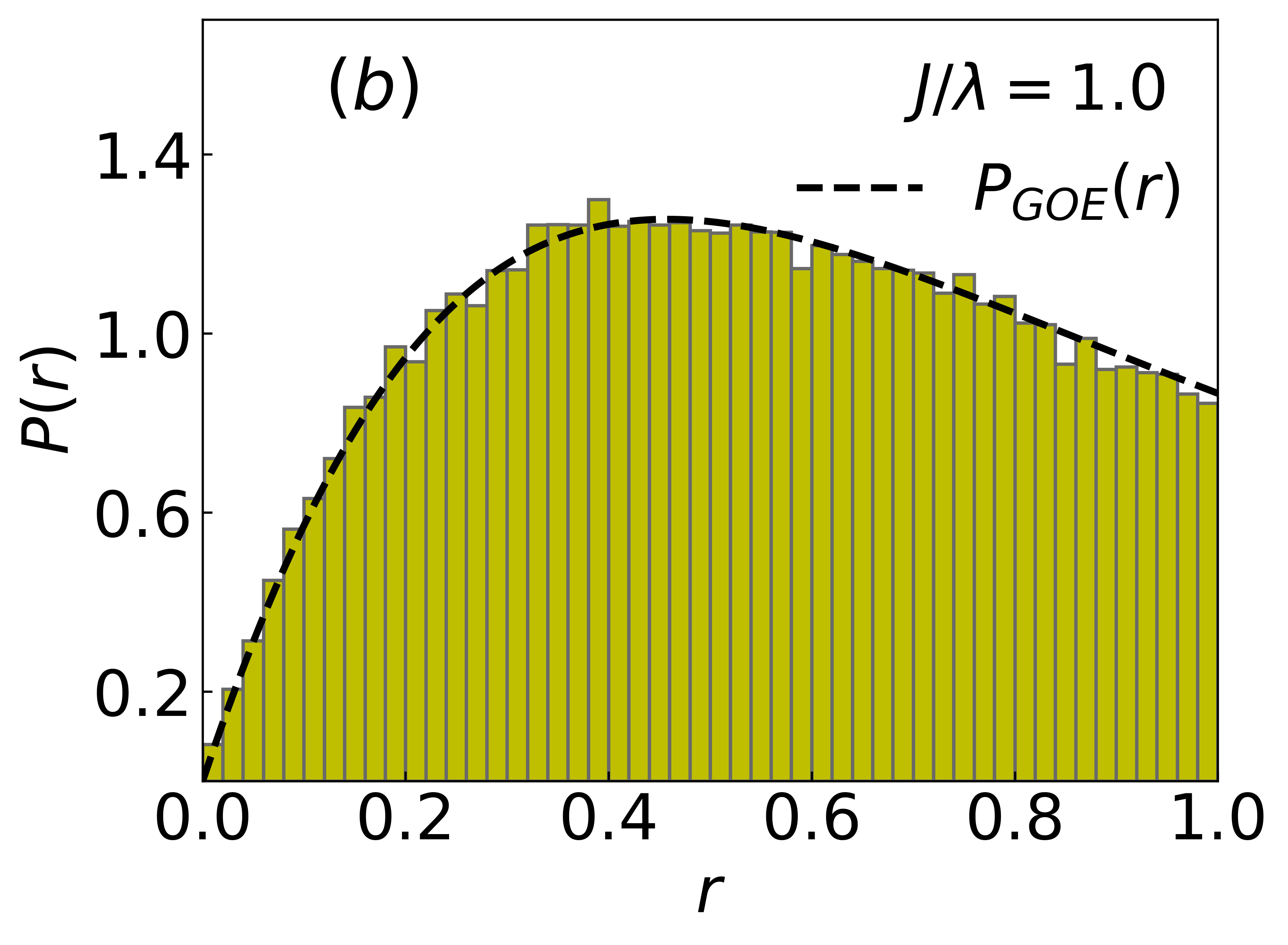} \\
	\includegraphics[scale=0.3]{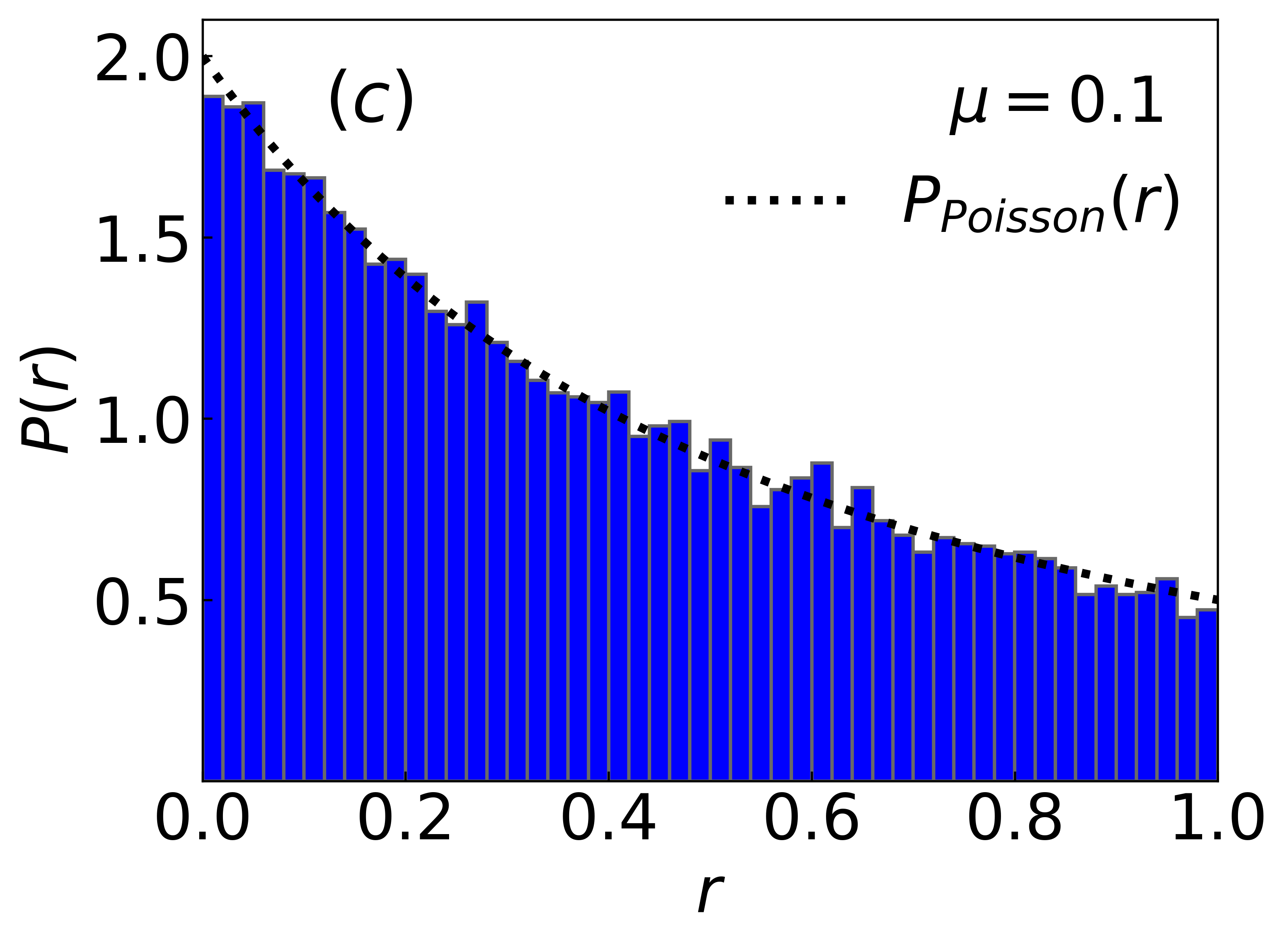}
	\includegraphics[scale=0.3]
	{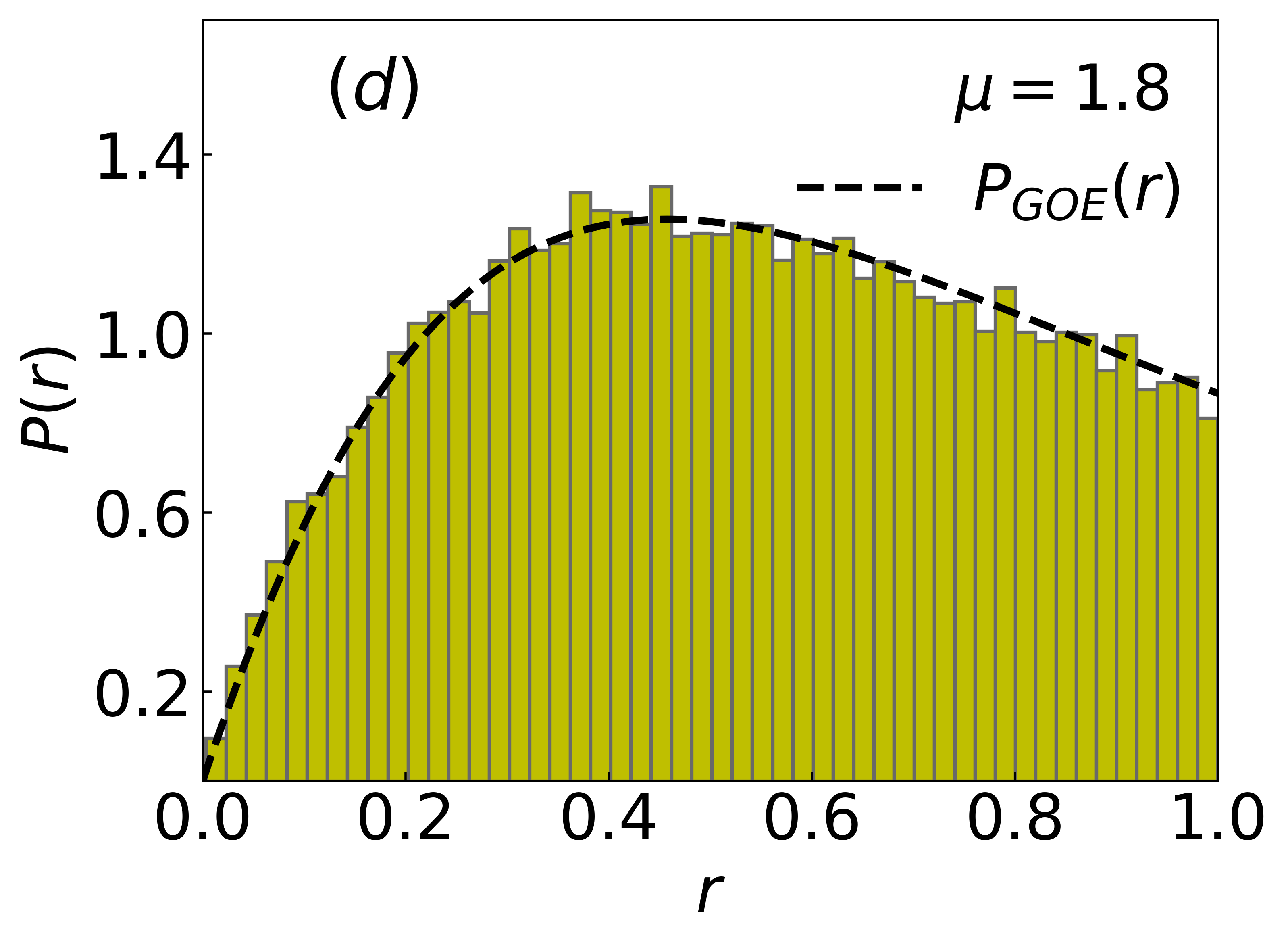}
	\caption{Distribution of adjacent gap ratios $P(r)$.
		(Top panel) Tavis-Cummings Lattice [(Eqs.~(1-3) in the main text] in the reflection-symmetric sector with $L=3, \ S=8$, and $N_{ex}=36$ for (a) weak hopping $J/\lambda=0.02$, and (b) strong hopping $J/\lambda=1.0$ .
		(Bottom panel) impurity model [Eq.~(4)] with $S = 64$ and $\lambda=1.0$  for (c) weak drive $\mu=0.1$, and (d) intermediate drive $\mu=1.8$. }
	\label{fig:P_r}
\end{figure}

\begin{figure}[htb!]
	\centering
	\includegraphics[scale=0.288]{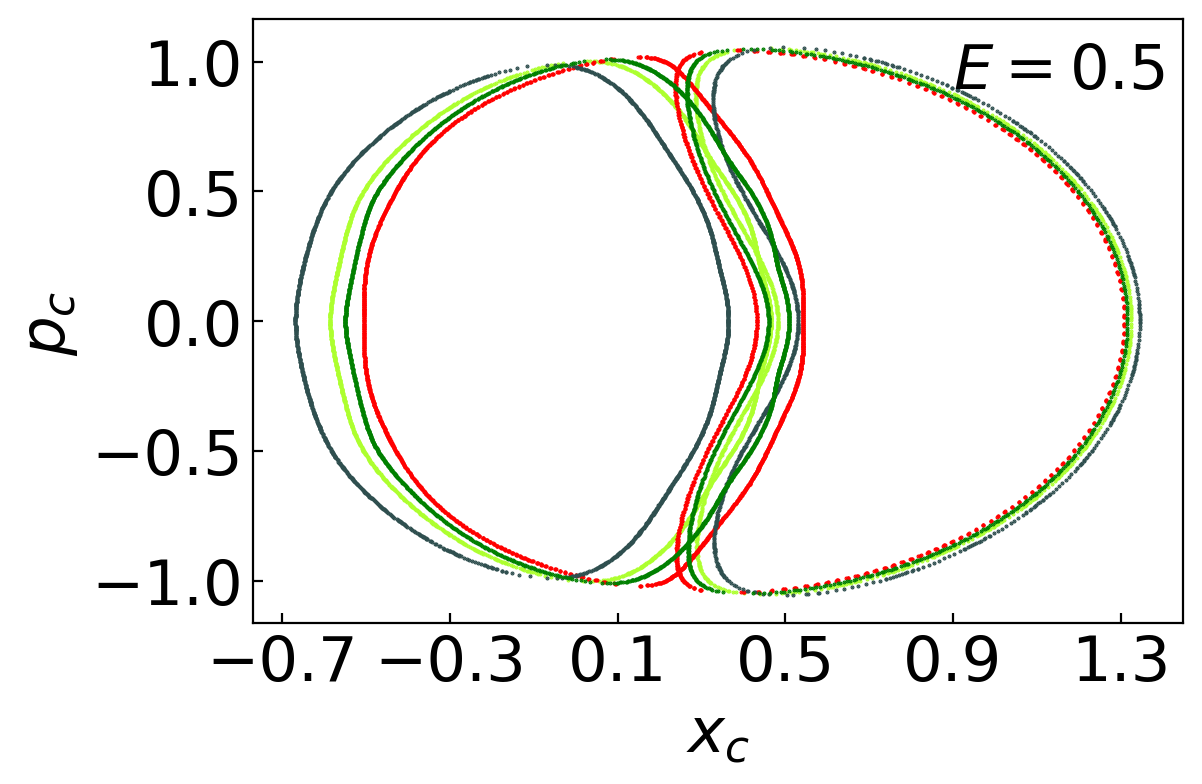}
	\includegraphics[scale=0.288]{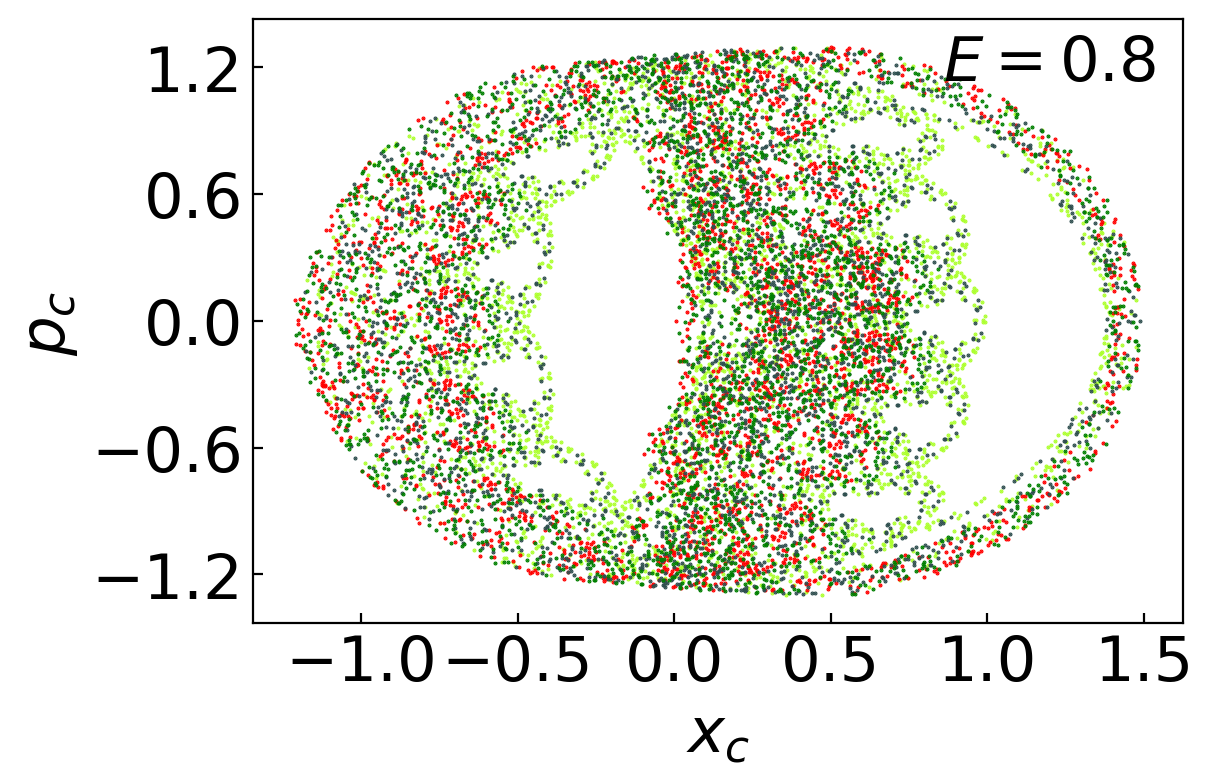}
	\includegraphics[scale=0.288]{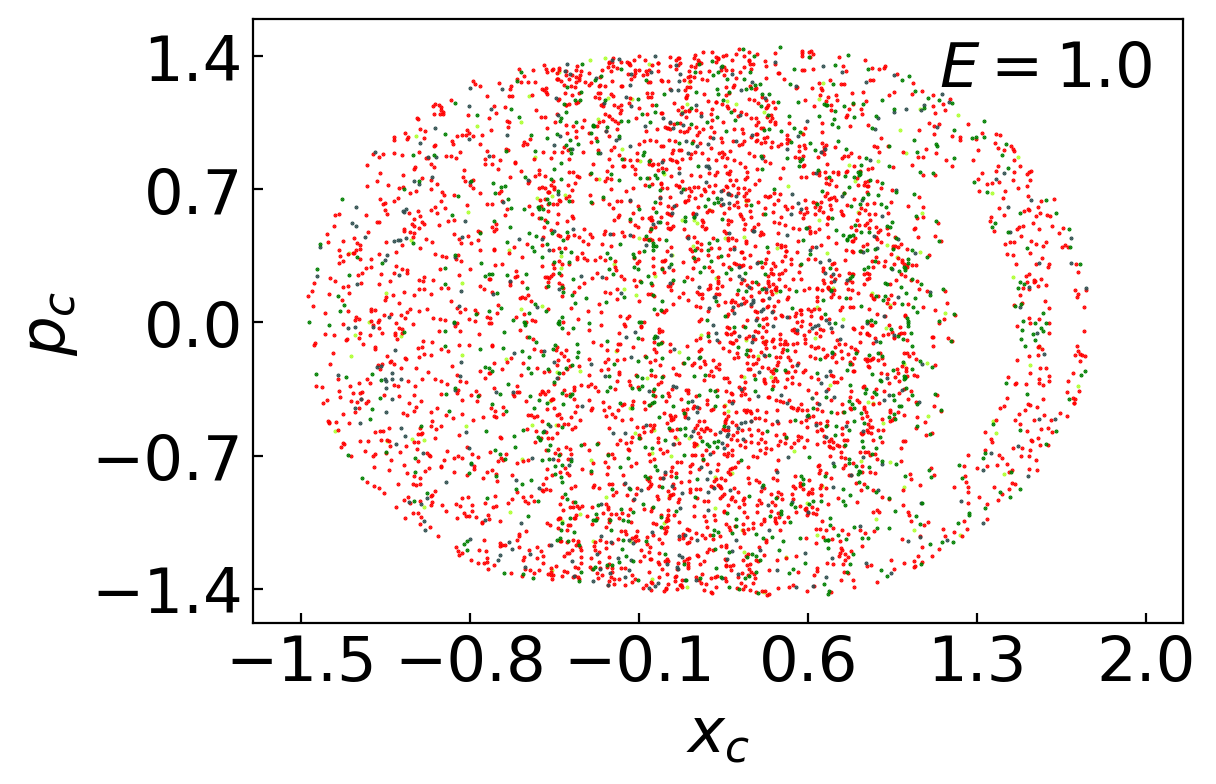}
	\includegraphics[scale=0.288]{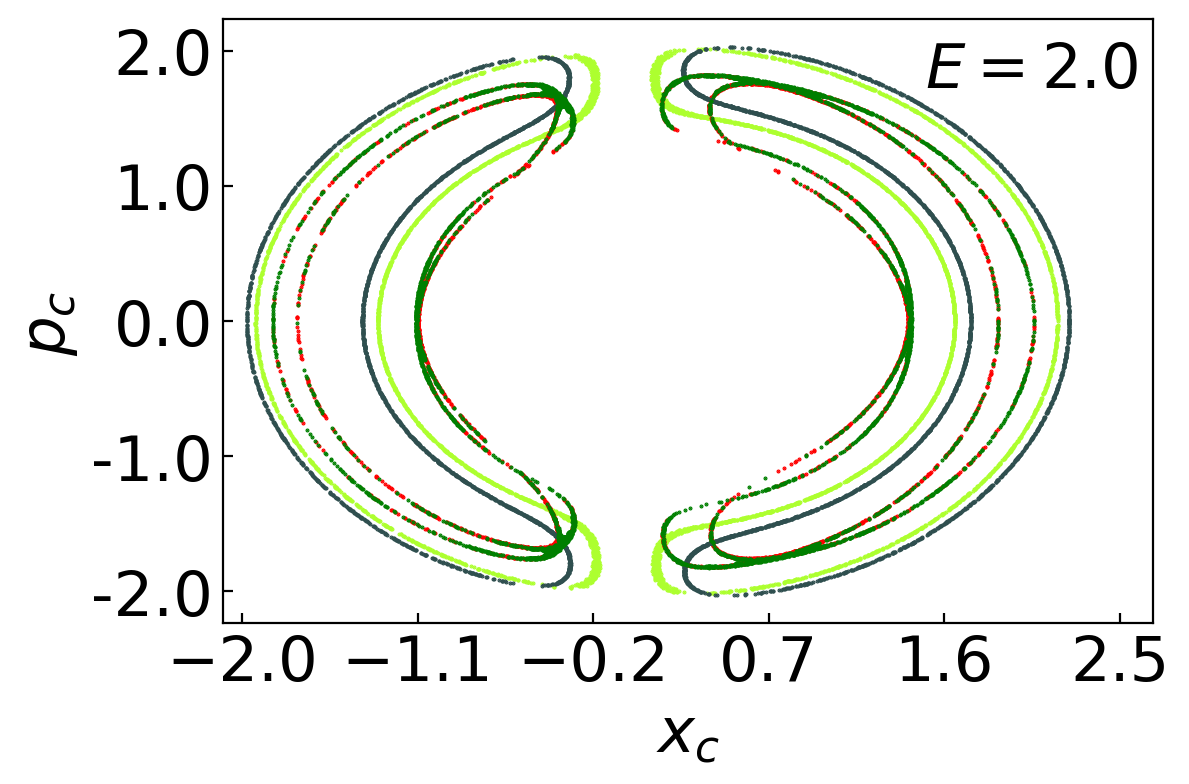}
	\includegraphics[scale=0.288]{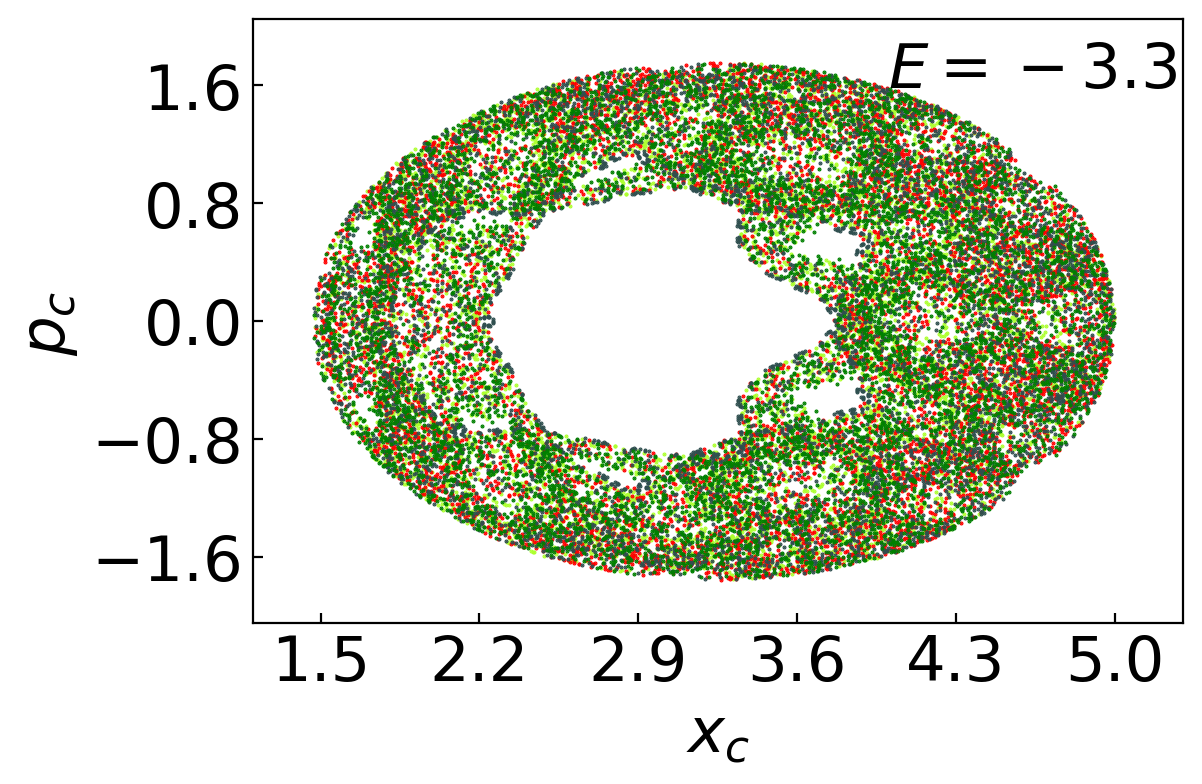}
	\includegraphics[scale=0.288]{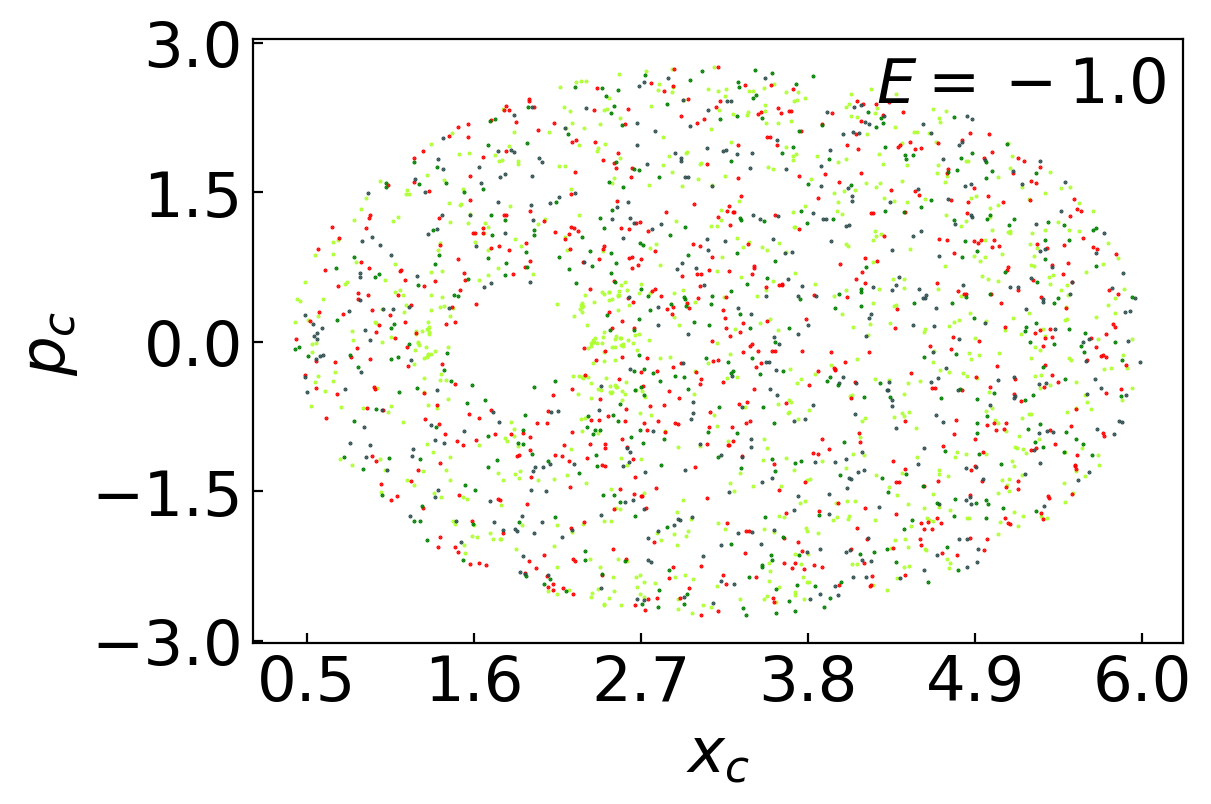}
	\includegraphics[scale=0.288]{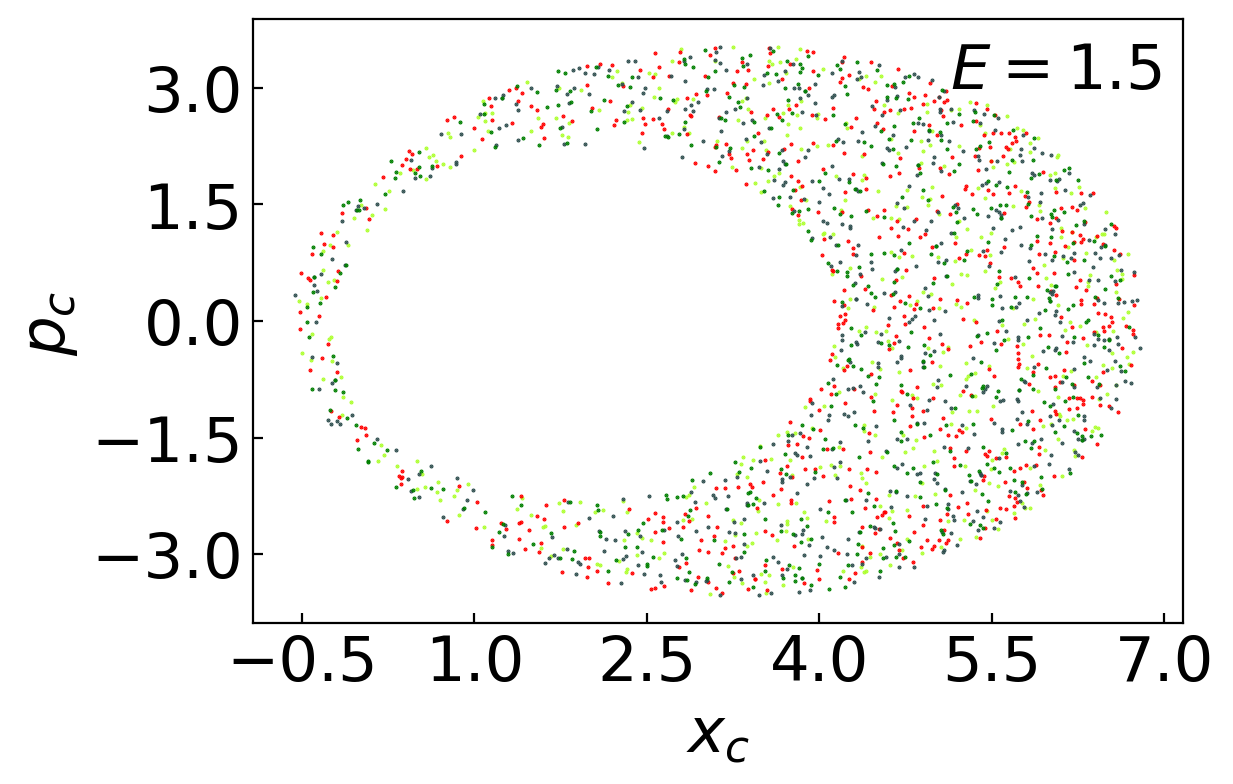}
	\includegraphics[scale=0.288]{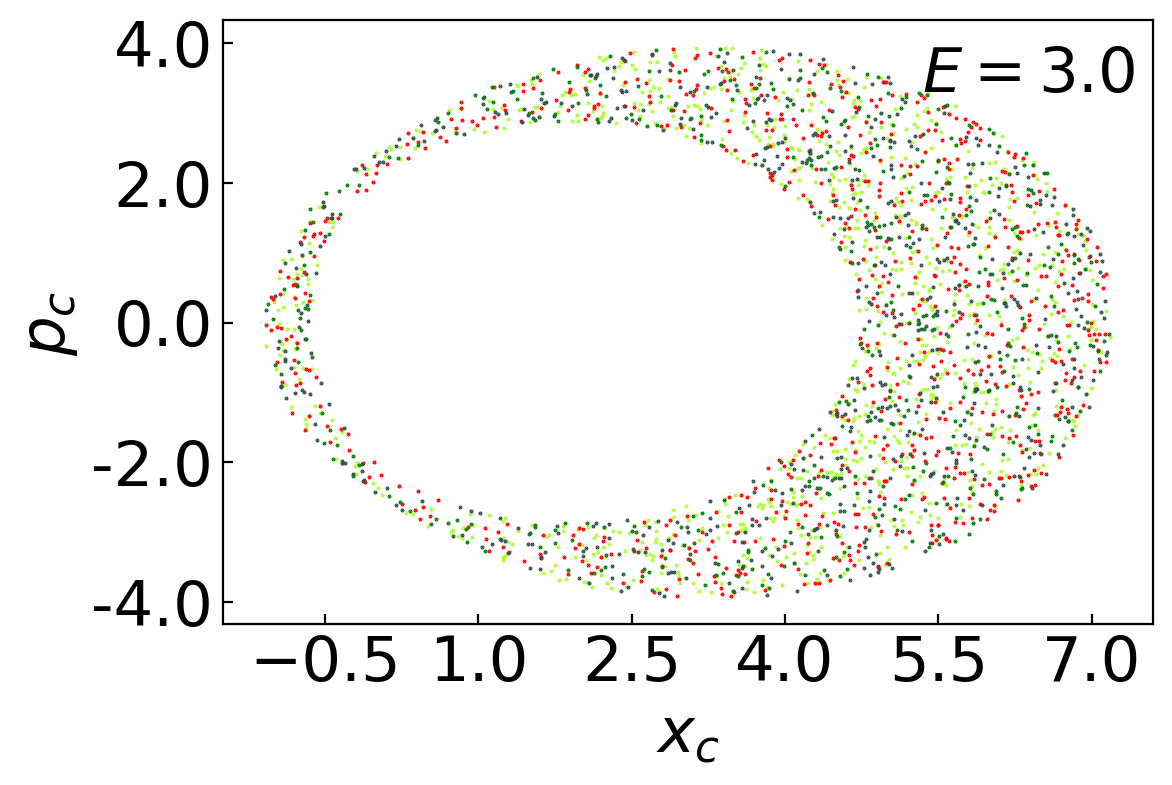}
	\includegraphics[scale=0.288]{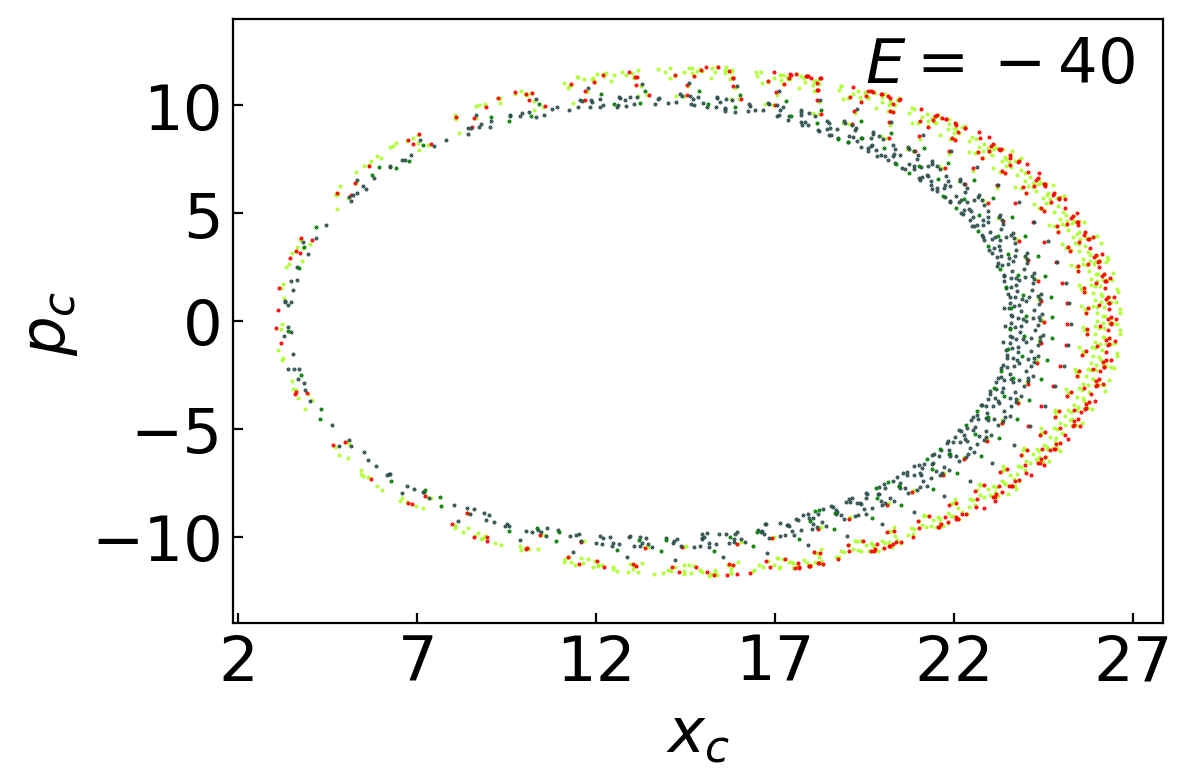}
	\includegraphics[scale=0.288]{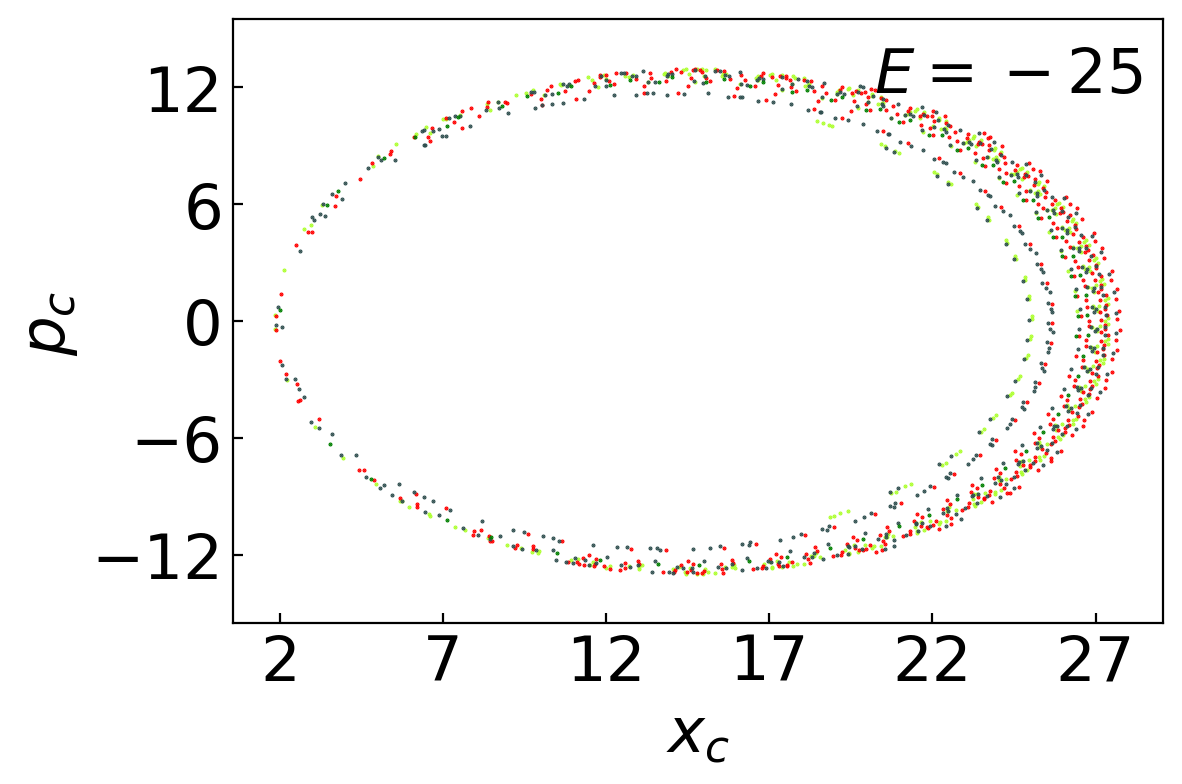}
	\includegraphics[scale=0.288]{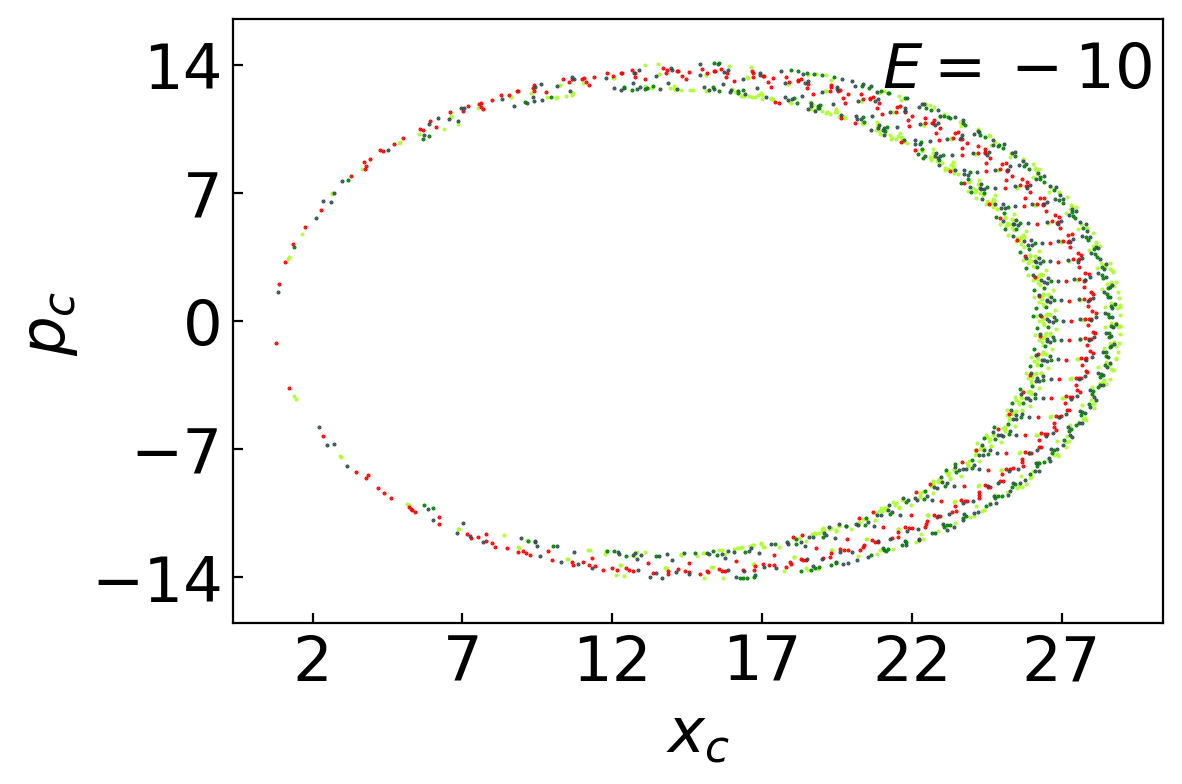}
	\includegraphics[scale=0.288]{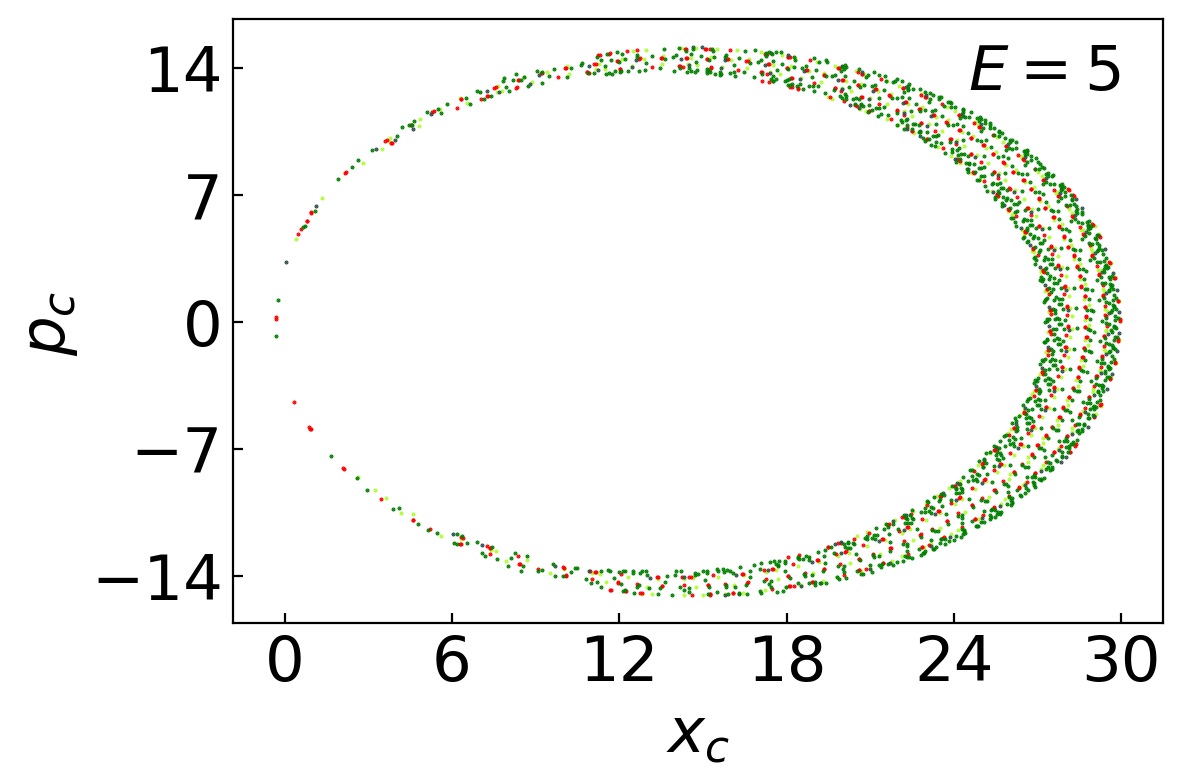}
	\caption{Poincar\'e sections of the classical driven impurity model defined in Eq.~(\ref{seq-c4}) in the limit $S\to \infty$, $\lambda=1.0$, and for various energies $E$ and (top) weak drive $\mu=0.1$, (middle) intermediate drive $1.8$, (bottom) strong drive $10.0$. Different colors represent trajectories with different initial conditions}
	\label{fig:poincare_section}
\end{figure}

\section{Chaos in the classical limit}
\label{classical}
In this section, we discuss the integrability to chaotic crossover of the classical limit of the dynamics of the impurity model~\cite{emary2003chaos2,corps2022chaos}.
We recall that the quantum Hamiltonian of the impurity model reads
\begin{align}
	H_{\rm imp} = & \, \omega_c \, a^{\dagger}a + \omega_s \left(S^{z}  +\frac{S}{2}\right) - \sqrt{S} \mu \left(a^\dagger + a\right) +\frac{\lambda}{\sqrt{S}} \left( a^{\dagger} S^- +  S^+ a  \right)\, . \label{seq-c1}
\end{align}
Note that we do not consider the dynamic scaling regime here.
The corresponding classical Hamiltonian is obtained in three steps: \\
(i) Express the spin operators in terms of bosonic operators using the Holstein-Primakoff transformation, yielding,
\begin{align}
	H_{\rm imp} =& \, \omega_c \, a^{\dagger}a + \omega_s\, b^\dagger b - \sqrt{S} \mu \left(a^\dagger + a\right) 
	+\frac{\lambda}{\sqrt{S}} \left( a^{\dagger}\sqrt{S - b^\dagger b} \, b +  
	b^\dagger \sqrt{S - b^\dagger b}\, a  \right) \,.  \label{seq-c2}
\end{align}
(ii) Writing the resulting Hamiltonian in terms of position and momentum operators defined as,
\begin{align}
	\hat{x}_c := \frac{1}{\sqrt{2 \omega_c}}\left(a^\dagger + a \right)\,, \
	\hat{p}_c := \rmi \sqrt{\frac{\omega_c}{2}}\left(a^\dagger - a \right)\,,\nonumber\\
	\hat{x}_s := \frac{1}{\sqrt{2 \omega_s}}\left(b^\dagger + b \right)\,, \
	\hat{p}_s := \rmi \sqrt{\frac{\omega_s}{2}}\left(b^\dagger - b \right)\,.
	\label{seq-c3}
\end{align}
(iii) Taking the classical limit, by replacing position and momentum operators by real numbers. This yields the classical Hamiltonian
\begin{align}
	H_{\rm imp}^{\rm cl} =& \frac{1}{2}\left(p_c^2+\omega_c^2 x_c^2 -\omega_c\right) + 
	\frac{1}{2}\left(p_s^2+\omega_s^2 x_s^2 -\omega_s\right)\nonumber\\
	&+ \lambda 
	\left(\sqrt{\omega_c \omega_s} x_c x_s + \frac{p_c p_s}{\sqrt{\omega_c 
			\omega_s}}\right) \eta(p_s,x_s) -\sqrt{S}\mu \sqrt{2 \omega_c} x_c,
	\label{seq-c4}
\end{align}
where $\eta(p_s,x_s):=\sqrt{1-\left(p_s^2+\omega_s^2 x_s^2 -\omega_s\right)/2 \omega_s S}$.

We consider the $S \to \infty$ limit by first rescaling the position and momentum coordinates as $\left(x_c,p_c,x_s,p_s\right) \mapsto \sqrt{S} \times \left(x_c,p_c,x_s,p_s\right)$, and the energies as $E \mapsto S \times E$. We obtain the following classical Hamilton's equations of motion~\cite{goldstein2002classical}
\begin{align}
	\frac{\rm d}{\rmd t} x_c = & \, p_c +\frac{\lambda}{\sqrt{\omega_c \omega_s}} \tilde\eta(x_s,p_s) p_s \,, \\
	\frac{\rmd}{\rmd t} p_c= &  - \omega_c ^2 x_c - \lambda \sqrt{\omega_c \omega_s} \tilde\eta(x_s,p_s) x_s + \sqrt{2 \omega_c} \mu  \,, \nonumber\\
	\frac{\rmd}{\rmd t} x_s= & \, p_s+\frac{\lambda}{\sqrt{\omega_c \omega_s}}  \Big[  \tilde\eta(x_s,p_s) p_c \nonumber \\
	& \qquad\qquad\ - \frac{1}{2\omega_s \tilde\eta(x_s,p_s) } \Big(\omega_c \omega_s \, x_c x_s+ p_c p_s \Big) p_s \Big]\,, \nonumber\\
	\frac{\rmd}{\rmd t} p_s = & - \omega_s ^2 x_s -\lambda \sqrt{\omega_c \omega_s} \Big[ \tilde\eta(x_s,p_s) x_c  \nonumber \\
	& \qquad\qquad\ - \frac{1}{2 \omega_c \tilde\eta(x_s,p_s)}  \Big(\omega_c \omega_s \, x_c x_s+ p_c p_s \Big) x_s \Big] \,,\nonumber
	\label{seq-c5}
\end{align}
where $\tilde{\eta}(x_s,p_s)=\sqrt{1-(p_s^2 +\omega_s ^2 x_s ^2)/2 \omega_s}$.
Note that the phase space is constrained by $0 \leq \tilde \eta(x_s,p_s) \leq 1$.
One simple and qualitative way to study the chaos in classical dynamics is to study Poincar\'e sections~\cite{goldstein2002classical,lakshmanan2012nonlinear}.
These are obtained by numerically integrating the above equations of motion, with the initial conditions set by
$x_c(0),p_c(0),x_s(0),p_s(0)$ and the energy $E$. 
The trajectories are projected on a chosen two-dimensional section.
We choose it to be the intersection between the hypersurface of  constant energy $E$, the hypersurface of equation $p_s(t)=0$, and the $(x_c,p_c)$ plane. 
Regular and structured Poincar\'e sections indicate integrable dynamics, whereas erratic and random structures indicate chaotic dynamics. 

The Poincar\'e sections generated for various initial conditions with different energies and for various values of the drive $\mu$ are displayed in Fig.~(\ref{fig:poincare_section}).
At small values of $\mu$ $(\mu=0.1)$, Poincar\'e sections are regular except for a small intermediate energy window where the dynamics are chaotic. This indicates a relative robustness of the $\mu=0$ integrable phase. If one views this phenomena through the lens of an energy-resolved version of Bohigas, Giannoni and Schmit (BGS) conjecture, it hints at the presence of extensive (in $S$) low-energy and high-energy portions of the spectrum of the \emph{quantum} impurity model whose universal features are dictated by Poisson statistics.

At intermediate values of the drive $\mu$ $(\mu=1.8)$, we observed chaotic dynamics at all the energies we numerically investigated. This hints at a quantum spectrum with statistical features dictated by random matrix theory.
Interestingly, at very large values of $\mu$ $(\mu=10)$, we observed close-to-integrable features at all energies. We attribute this to a drive term which is so strong that it effectively screens the effect of the non-linearity $\lambda$ that is responsible for chaos. The same reasoning can be applied to the quantum version of the model.

{
\section{Mean-field relation between lattice and impurity spectral form factors}
\label{lattice-impurity}
In this Appendix, we use a mean-field approach to motivate the impurity model presented in the manuscript. The mean-field approximation is a standard approach to compute thermodynamics but, here, we perform it in the setting of computing the spectral form factor (SFF). 
We work in the context of the Tavis-Cummings Hamiltonian on a ``mean-field lattice'' with all-to-all couplings between the $L$ sites, where $L$ is large,
\begin{eqnarray} \label{eq:all2all}
 H &= & \sum_{i=1}^{L} h_i - \frac{J}{L} \sum_{i,j=1}^{L} a_i^\dagger a_j\, ,\\
h_{i} &= & \, \omega_c a_i^{\dagger}a_i + \omega_s S_i^{z}
+\frac{\lambda}{\sqrt{S}} \left( a_i^{\dagger} S_i^- +  a_i S_i^+  \right) \,.
\end{eqnarray}
We have scaled the hopping term by $1/L$, which is the standard convention consistent with a non-trivial thermodynamic limit. The above Hamiltonian is $U(1)$-symmetric which corresponds to the conservation law $\left[H,N\right]=0$ with $N=\sum_{i=1}^{L}n_i$ where $n_i:=a_i^\dagger a_i + S_i^z+\frac{S}{2}$.
Additionally, the Hamiltonian is symmetric under permutations of the sites. This extra symmetry, artifact of the all-to-all geometry, introduces a subtlety in that the SFF which is relevant for diagnosing quantum chaotic features should in principle be computed in a given sector of the permutation group.
A proper treatment of this permutation symmetry is technically challenging and we postpone this computation to future work. Here below we adopt the following strategy:
\begin{itemize}
\item We first present a detailed computation that avoids this subtlety by explicitly breaking the permutation symmetry, therefore computing the mean-field SFF from the untruncated spectrum (yet properly accounting for the $U(1)$ symmetry). This will yield a simple relation between the lattice and impurity SFFs, see Eq.~(\ref{eq:KiskL}).
\item Secondly, we present a speculative discussion on how to modify the results of the former computation to account for the permutation symmetry group and we speculate the simple relation~(\ref{eq:graal}) between the lattice and impurity SFFs.
\end{itemize}

\paragraph{Explicitly breaking permutation symmetry.}
Let us first avoid the technicalities related to the permutation group by explicitly breaking this extra symmetry. This can be done by adding disorder with random onsite energy shifts: $h_i \to h_i + \delta_i \left(a_i^\dagger a_i + S_i^z\right)$ where $\delta_i$ is random-valued. We shall later take $\delta_i \to 0$.

Working in the sector with $N= L\rho$ particles, where $\rho$ is the particle density, we define the corresponding SFF as
\begin{eqnarray}
 K_{L}^{\rho}(t) &:=& \langle|\tilde{Z}_{L}^{}(t,\rho)|^2\rangle, \ t \geq 0\,,
\end{eqnarray}
where $\langle \cdots \rangle$ correspond to disorder averaging and $\tilde{Z}_{L}^{}(t,\rho)$ is the imaginary-temperature partition function of the lattice model in the $N = L \rho$ particle sector and for a given realization of the disorder.
It is defined as,
\begin{eqnarray}
\label{project}
\tilde{Z}_{L}^{}(t,\rho) &:=& {\rm Tr}\left[\rme^{-\rmi H t} \delta(L \rho - N)\right]\,,
\end{eqnarray}
where ${\rm Tr}$ is the trace over the full lattice Hilbert space and the delta function implements the partial trace on those states with exactly $N=L\rho$ particles. It can be expressed in terms of an imaginary-temperature grand-canonical partition function via
\begin{eqnarray}
\label{eq:buffer1}
 \tilde{Z}_{L}^{}(t,\rho) &=& \int_{-\infty}^{+\infty} \frac{\rmd \nu}{2\pi}\rme^{-\rmi L \rho \nu} Z_{L}^{}(t,\nu)\,,
\end{eqnarray}
where $\nu$ plays the role of a chemical potential and the grand-canonical lattice partition function is defined as
\begin{eqnarray}
 Z_{L}^{}(t,\nu) &:=&  {\rm Tr}\left[\rme^{-\rmi\left[H-\frac{\nu}{t} N\right]t}\right]\,.
\end{eqnarray}

We now relate the partition function of the lattice to that of impurity by employing a standard Hubbard-Stratonovich decoupling of the hopping term in the Hamiltonian, and a subsequent saddle-point approximation. The latter is sometimes dubbed  the strong-coupling random phase approximation (RPA) in the Bose-Einstein condensation literature. The saddle-point approximation will become exact in $L \to \infty$ limit.

Introducing complex Hubbard-Stratonovich fields $\Phi(\tau)$ and $\Phi^*(\tau)$ for $\tau \in [0,t]$, $Z_{L}(t,\nu)$ can be rewritten as
\begin{eqnarray}
\label{hst}
Z_{L}(t,\nu) 
&=& \int_{\Phi(t)=\Phi(0)}  \hspace{-2em }\mathcal{D}[\Phi,\Phi^*] \, \rme^{\rmi J L \int_{0}^{t}\rmd\, \tau \Phi^*(\tau)\Phi(\tau)}\nonumber\\
&& \qquad \times {\rm Tr}\left[\overleftarrow{\mathcal{T}}\rme^{-\rmi\int_{0}^{t} \rmd\tau \, \sum_{i=1}^{L}\left[h_i-\frac{\nu}{t} n_i + J \left(\Phi^*(\tau) a_i + a_i^\dagger \Phi(\tau)\right)\right]}\right]\nonumber\\
&=&
\int_{\Phi(t)=\Phi(0)}  \hspace{-2em}
 \mathcal{D}[\Phi,\Phi^*] \,\rme^{\rmi J L \int_{0}^{t}\rmd\tau \, \Phi^*(\tau)\Phi(\tau)} \prod_{i=1}^{L} \mathfrak{z}_{i}(t,\nu;[\Phi,\Phi^*])\,,
\end{eqnarray}
where $\mathcal{D}[\Phi,\Phi^*]$ is the functional integral measure over complex functions on $[0,t]$. $\overleftarrow{\mathcal{T}}$ is the time-ordering operator. $\mathfrak{z}_i$ is the partition function of a local impurity model, namely the single site $i$ in the presence of auxiliary Hubbard-Stratonovich fields which act as external drives. It reads
\begin{eqnarray}
 \mathfrak{z}_i(t,\nu,[\Phi,\Phi^*])&:=&{\rm tr}\left[\overleftarrow{\mathcal{T}}\rme^{-\rmi\int_{0}^{t} \rmd\tau \left[  h_i - \frac{\nu}{t} n + J \left(\bar{\Phi}^*(\tau) a + a^\dagger \bar{\Phi}(\tau) \right) \right]}\right]\,.
\end{eqnarray}
where ${\rm tr}$ is the trace in the local Hilbert space at site $i$.
Sending the random onsite energy shifts to zero, $\delta_i \to 0$, we now get $ \mathfrak{z}_i \to \mathfrak{z}$ for all sites with
\begin{eqnarray}
\mathfrak{z}(t,\nu,[\Phi,\Phi^*]) :=
 {\rm tr}\left[\overleftarrow{\mathcal{T}}\rme^{-\rmi\int_{0}^{t} \rmd \tau \left[ h-\frac{\nu}{t} n + J \left(\bar{\Phi}^*(\tau) a + a^\dagger \bar{\Phi}(\tau) \right)  \right]} \right]\,.
\end{eqnarray}
We have
\begin{eqnarray}
Z_{L}(t,\nu) &=&\int_{\Phi(t)=\Phi(0)}  \hspace{-2em}
 \mathcal{D}[\Phi,\Phi^*] \, \rme^{\rmi L \left[J \int_{0}^{t}\rmd\tau \, \Phi^*(\tau)\Phi(\tau)-\rmi\ln\mathfrak{z}(t,\nu,[\Phi,\Phi^*])\right]}\,. \label{eq:D10}
\end{eqnarray}
We now estimate the above path integral by a saddle-point approximation, \textit{i.e.} by extremizing the action over $\Phi(\tau)$, which becomes exact in $L \to \infty$ limit. Restricting to time-independent solutions $\bar \Phi$ and $\bar \Phi^*$, the latter are governed by self-consistent equations
\begin{eqnarray}
 \bar{\Phi}= \frac{{\rm tr}\left[a \, \rme^{-\rmi\, h_{\rm imp}(t,\nu,\bar\Phi,\bar\Phi^*)\, t}\right]}{\mathfrak{z}(t,\nu,\bar{\Phi},\bar{\Phi}^*)}, \quad
 \bar{\Phi}^*= \frac{{\rm tr}\left[a^\dagger \, \rme^{-\rmi \, h_{\rm imp}(t,\nu,\bar\Phi,\bar\Phi^*)\, t }\right]}{\mathfrak{z}(t,\nu,\bar{\Phi},\bar{\Phi}^*)}\,, \label{eq:buffer3a}
\end{eqnarray}
with the impurity partition function
\begin{eqnarray}
\mathfrak{z}(t,\nu,\Phi, \Phi^*) =  {\rm tr}\left[\rme^{-\rmi \, h_{\rm imp}(t,\nu,\Phi,\Phi^*)\, t }\right]\,,
\end{eqnarray}
and where we have introduced the impurity Hamiltonian
\begin{eqnarray} \label{eq:impurity_app}
h_{\rm imp}(t,\nu,\Phi,\Phi^*) := h-\frac{\nu}{t} n + J \left(\Phi^* a + a^\dagger \Phi \right)\,.
\end{eqnarray}
We can now express the lattice partition function in terms of the impurity partition function as
\begin{eqnarray}
\label{eq:buffer2}
Z_{L}(t,\nu) &\overset{L \to \infty}{\approx}& \rme^{\rmi L J \bar{\Phi}^* \bar{\Phi} t} \left[\mathfrak{z}(t,\nu,\bar{\Phi},\bar{\Phi}^*)\right]^L\,.
\end{eqnarray}
Using Eq. (\ref{eq:buffer2}) in Eq. (\ref{eq:buffer1}), and evaluating $\nu$ integral by means of another saddle-point approximation, we obtain
\begin{eqnarray}
 \tilde{Z}_{L}(t,\rho) &\overset{L \to \infty}{\approx}& \rme^{-\rmi L \rho \bar{\nu}+ \rmi J L \bar{\Phi}^* \bar{\Phi} t} \left[\mathfrak{z}(t,\bar{\nu},\bar{\Phi}, \bar{\Phi}^*)\right]^L\,,
\end{eqnarray}
where the saddle-point value of the chemical potential $\bar{\nu}$ is governed by the equation
\begin{eqnarray}
 \label{eq:buffer4}
 \rho &=& \frac{{\rm tr}\left[n \, \rme^{-\rmi t \left[h-\frac{\bar{\nu}}{t} n + J \left(\bar{\Phi}^* a + a^\dagger \bar{\Phi}\right)\right]}\right]}{\mathfrak{z}(t,\bar{\nu},\bar{\Phi}, \bar{\Phi}^*)}\,.
\end{eqnarray}

The SFF of the lattice model can now be expressed in terms of that of the impurity model as,
\begin{eqnarray}
\label{latimp}
 K_{L}^{\rho}(t) &\overset{L \to \infty}{\approx}& |\rme^{-\rmi \rho \bar{\nu}+\rmi  J \bar{\Phi}^* \bar{\Phi} t}|^{2L} k(t)^L\,,
\end{eqnarray}
where $k(t)$ is the impurity SFF, defined as
\begin{eqnarray}
 k(t)&:=& |\mathfrak{z}(t,\bar{\nu},\bar{\Phi}, \bar{\Phi}^*)|^2\,,
\end{eqnarray}
and $\bar{\Phi}$, $\bar{\Phi}^*$ and $\bar\nu$ are determined from the saddle-point equations Eqs.~(\ref{eq:buffer3a}) and~(\ref{eq:buffer4}).
Note that hermiticity of the impurity Hamiltonian at the saddle point requires $\bar \nu$ to be real and $\bar\Phi$ and $\bar\Phi^*$ to be complex conjugate of each other.
Under those assumptions, we obtain
\begin{eqnarray} \label{eq:KiskL}
 K_{L}^{\rho}(t) &\overset{L \to \infty}{\approx}& k(t)^L\,,
\end{eqnarray}
where $ K_{L}^{\rho}(t)$ is the SFF of the all-to-all lattice model defined in Eq.~(\ref{eq:all2all}) and $k(t)$ is the one of the impurity model defined in Eq.~(\ref{eq:impurity_app}).
The above equality has to be understood as a mean-field factorization of the lattice SFF into a product of those of the local impurity sites. This result is naturally consistent with the general expectation that a mean-field approach formally decouples the lattice partition function in terms of a product of local partition functions.
Equation~(\ref{eq:KiskL}) is trivially obeyed at $t\to \infty$ and at $t=0$. Indeed, in the former case, the SFFs simply reduce to their respective Hilbert space dimensions, and they reduce to their square in the latter case.

\paragraph{Speculation in the totally-symmetric sector.}
Computing the lattice SFF constrained to the totally-symmetric sector of the permutation group relies on estimating the partition function
\begin{eqnarray}
\tilde{Z}_{L}^{}(t,\rho) &:=& {\rm Tr}\left[\rme^{-\rmi H t} \delta(L \rho - N)\mathbb{P}_{S}\right]\,,
\end{eqnarray}
where $\mathbb{P}_{S}$ is the projector onto the totally-symmetric subspace.
Let us now present arguments leading to an educated guess for this lattice SFF.
The idea is to start from the factorized result in Eq.~(\ref{eq:KiskL}) and rework it in order to see it as resulting from a single trace of a permutation-symmetric operator. 

Let us first introduce the eigen-basis of the impurity Hamiltonian $h_{\rm imp}$ evaluated at the saddle point: $h_{\rm imp}|e_n \rangle = e_n |e_n \rangle $ for $n=1\ldots d$ where $d$ is the dimension of the local Hilbert space.
Rewriting the {\sc rhs} of Eq.~(\ref{eq:KiskL}) as $k(t)^L =   k(t) \times k(t) \times \ldots  \times  k(t)= |\sum_{n_1,\cdots,n_L}\langle e_{n_1}| \rme^{-\rmi\, h_{\rm imp}\, t}|e_{n_1}\rangle\cdots\langle e_{n_L}| \rme^{-\rmi \, h_{\rm imp}\, t}|e_{n_L}\rangle|^2$, and given that, in the mean-field approach, the relevant part of the lattice Hilbert space is composed of factorized states of the form $|e_{n},\cdots,e_n \rangle := |e_n\rangle^{(1)}\ldots|e_n\rangle^{(L)} $, we propose the following expression for the lattice SFF in the permutation-symmetric sector at large $L$: $K^\rho_L(t)  \stackrel{?}{\approx} |\sum_{n}\langle e_{n}| \rme^{-\rmi h_{\rm imp} t}|e_{n}\rangle\cdots\langle e_{n}| \rme^{-\rmi \, h_{\rm imp}\, t}|e_{n}\rangle|^2 = |\sum_{n}\langle e_{n},\cdots,e_n| \rme^{-\rmi (h_{\rm imp}^{(1)}+\cdots+h_{\rm imp}^{(L)}) t}|e_{n},\cdots,e_n\rangle|^2$ where the latter expression clearly displays the permutation symmetry. This amounts in speculating that in the permutation-symmetric sector, and at large $L$, the lattice SFF is related to the impurity SFF via 
\begin{eqnarray}
K^\rho_L(t) \stackrel{?}{\approx}_{L\to\infty}  k(L t)
\end{eqnarray}
rather than the relation (\ref{eq:KiskL}) which was derived without consideration for permutation symmetry. In the {\sc rhs}, $L$ enters as a multiplicative scale to the impurity spectrum. However, this scale is simply gauged out if one computes the SFF from unfolded spectra as is customary (see~\ref{unfold}). Ultimately, this amounts in speculating the simple relation
\begin{eqnarray}\label{eq:graal}
K^\rho_L(t) \stackrel{?}{\approx}  k(t)\,,
\end{eqnarray}
where $K^\rho_L(t)$ is the SFF computed from the totally-symmetric sector of the unfolded spectrum of the all-to-all lattice model in defined in Eq.~(\ref{eq:all2all}) and $k(t)$ is the SFF of the unfolded impurity model defined in Eq.~(\ref{eq:impurity_app}).
}

\bigskip

\section*{References}
\bibliography{references.bib}

\end{document}